\documentclass[preprint,prl,showpacs,superscriptaddress]{revtex4-1} 

\usepackage{amssymb}
\usepackage{subfigure}
\usepackage{amsmath}

\usepackage{graphicx}
\usepackage{epstopdf}
\usepackage{color}
\usepackage{ulem}

\newcommand{\scl}{0.2} 

\begin{document}


\title{Spectral Signatures of Exceptional Points and Bifurcations\\
in the Fundamental Active Photonic Dimer}

\date{\today}

\author{Yannis Kominis}
\affiliation{School of Applied Mathematical and Physical Science, National Technical University of Athens, Athens, Greece	}

\author{Vassilios Kovanis}
\affiliation{Department of Physics, School of Science and Technology, Nazarbayev University, Astana, Republic of Kazakhstan}
\affiliation{CREOL/College of Optics and Photonics, University of Central Florida, Orlando, Florida, USA}

\author{Tassos Bountis}
\affiliation{Department of Mathematics, School of Science and Technology, Nazarbayev University, Astana, Republic of Kazakhstan}

\begin{abstract}
The fundamental active photonic dimer consisting of two coupled quantum well lasers is investigated in the context of the rate equation model. Spectral transition properties and exceptional points are shown to occur under general conditions, not restricted by PT-symmetry as in coupled mode models, suggesting a paradigm shift in the field of non-Hermitian photonics. The optical spectral signatures of system bifurcations and exceptional points are manifested in terms of self-termination effects and observable drastic variations of the spectral line shape that can be controlled in terms of optical detuning and inhomogeneous pumping.
\end{abstract}

\maketitle

Non-Hermitian photonics is a research field of continuously increasing interest from a theoretical and technological point of view. Non-Hermitian systems are known to have unique mathematical properties that have no counterpart in Hermitian ones and are directly related to features of great potential for technological applications in integrated photonics and active metasurfaces. Non-Hermitian photonic systems are characterized by configurations with inhomogeneous distribution of gain and loss. The fundamental non-Hermitian photonic element is a dimer consisting of two coupled lasers operating under gain and loss conditions, respectively, which has served as the basic paradigm for the study of non-Hermitian optics under PT-symmetry conditions \cite{PT_1, PT_2, PT_3, PT_4, Ramezani_2010, Longhi_2015}.\

The commonly used underlying model consists of a system of coupled mode equations with fixed gain and loss, and it has been shown that under PT-symmetry conditions, corresponding to zero detuning between the cavities and exact balance between gain and loss, the system supports an interesting set of features such as spectral transitions and existence of exceptional points \cite{Heiss}. Exceptional points occur naturally in eigenvalue problems that depend on parameters. Varying such parameters, one can generically find points where eigenvalues coincide and the corresponding eigenvectors collapse. Recently there has been a strong interest in the physics and potential applications related to single-mode lasing \cite{Christodoulides_2014, Feng_2014, Christodoulides_2015, Coldren_2016}, reverse pump dependence of the laser power and self-termination \cite{Rotter_2012, Rotter_2014, Peng_2014, El-Ganainy_2014}, enhanced laser sensitivity \cite{Khajavikhan_2017, CREOL_2017}, chiral properties of metasurfaces \cite{KangChenChong} and topological effects in laser arrays \cite{Parto_2017, El-Ganainy_2017}. 
Deviations from PT-symmetry conditions have been studied \cite{Christodoulides_2015} and it has been shown that for linear systems, the essential condition for the existence of the these interesting spectral features is the zero detuning condition \cite{Choquette_2017, Abouraddy_2017}. The effect of nonlinearity on PT-symmetric systems has been studied \cite{Segev_2013, Ge_2016, Hassan_2017} and it has been shown that in the completely asymmetric active dimer, nonlinear supermodes of finite power exist \cite{Kominis_2016} and are modulationally stable \cite{Kominis_2017}.\

However, the aforementioned simplified model neglects significant dynamical features of the system such as the nonlinear coupling between the electric field and the carrier density dynamics as well as the amplitude-to-phase coupling. In this work, we study the properties of the fundamental active photonic dimer, by utilizing the realistic rate equation model that incorporates these important effects \cite{Wang_88, Erneux_book}. It is shown that spectral transitions and existence of exceptional points occur under more general conditions than those implied by PT-symmetry, including the case of nonzero detuning. Moreover, it is shown that the spectral properties of the system are directly controlled by the pumping rates of the two lasers. The generalization of the conditions for the existence of interesting spectral features along with the controllability of the system suggests a paradigm shift in the study of non-Hermitian photonics and facilitates a large variety of experiments and applications. We show that the linear eigenvalue spectrum of the zero state solution can directly explain reverse pump dependence of the laser output and self-termination in terms of bifurcations of the zero state stability, either with or without the presence of exceptional points, under general conditions. Moreover, the model is capable of providing the linear eigenvalue spectrum of the non-zero phase-locked states of the system, that are the actual operating modes of the dimer. It is shown that bifurcations and exceptional points directly manifest themselves in experimentally observable characteristics of the dimer in terms of intensity peaks and sideband branching in the spectral line shape of the output of the system. \    

The dynamics of two evanescently coupled semiconductor lasers is governed by the following equations for the slowly varying complex amplitude of the normalized electric field $\mathcal{E}_i$ and the normalized excess carrier density $N_i$ of each laser:
\begin{eqnarray}
 \frac{d\mathcal{E}_i}{dt}&=&(1-i\alpha)\mathcal{E}_i N_i+i\eta\mathcal{E}_{3-i} +i\omega_i \mathcal{E}_i \nonumber \\
 T\frac{dN_i}{dt}&=&P_i-N_i-(1+2N_i)|\mathcal{E}_i|^2 \hspace{3em} i=1,2 \label{array}
\end{eqnarray}
where $\alpha$ is the linewidth enhancement factor, $\eta$ is the normalized coupling constant, $P_i$ is the normalized excess electrical pumping rate (positive and negative values correspond to pumping above and below the lasing threshold for a single laser), $\omega_i$ is the normalized optical frequency detuning from a common reference, $T$ is the ratio of carrier to photon lifetimes, and $t$ is the normalized time \cite{Wang_88}. 
The uncoupled lasers ($\eta=0$), exhibit free running relaxation with frequencies $ \Omega_i=\sqrt{2P_i/T}$. In the following, we use a reference value $P$ in order to define a frequency $ \Omega=\sqrt{2P/T}$ which is untilized for the following rescaling: $Z_i \equiv N_i/\Omega, \Lambda\equiv\eta/\Omega, \Omega_i\equiv\omega_i/\Omega$ and $\tau\equiv\Omega t$. As a reference case, we consider a pair of lasers with $\alpha=5$, $T=400$ which is a typical configuration relevant to experiments and we take $P=0.5$. For these values we have $\Omega=5 \times 10^{-2}$ and a coupling constant $\eta$ in the range of $10^{-5} \div 10^0$ corresponds to a $\Lambda$ in the range $0.5 \times 10^{-3} \div 0.5 \times 10^{2}$.\


Firstly, we investigate the spectral properties of the zero-state of the system (\ref{array}). By linearizing the system around the zero state solution 
$ \mathcal{E}_i=0, Z_i=P_i/\Omega (i=1,2)$  we obtail a linear system with eigenvalues obtained by the characteristic equation $ \det(A-\lambda I)  \det(A^*-\lambda I)\det(B-\lambda I)=0$ with 
\begin{equation}
 A=\begin{bmatrix}
    (1-i\alpha)P_1/\Omega + i \Omega_1      &  i \Lambda  \\
    i \Lambda      & (1-i\alpha)P_2/\Omega +i \Omega_2 \\
   \end{bmatrix},
\end{equation}
and $B=-\Omega/(2P) I_{2\times2}$, where $I_{2\times2}$ is the $2\times2$ unit matrix. The eigenvalues are given by 
\begin{equation}
 \lambda_{1,2}=\Lambda \left \{ \bar{P}+i\left(\bar{\Omega}-\alpha\bar{P}\right) \pm \sqrt{ \left[\Delta P+i\left(\Delta-\alpha \Delta P\right)\right]^2-1}\right \},
\end{equation}
$\lambda_{3,4}=\lambda_{1,2}^*$, and  $\lambda_{5,6}=-1/(2P)$, where $\bar{P}=(P_1+P_2)/2\Lambda$, $\bar{\Omega}=(\Omega_1+\Omega_2)/2\Lambda$, $\Delta P=(P_1-P_2)/2\Lambda$, $\Delta=(\Omega_1-\Omega_2)/2\Lambda$ and, without loss of generality, we have set $\Omega=1$. It is clear that $\lambda_{5,6}$ are always negative whereas the eigenvalues $\lambda_{1,2}$ dictate the stability of the zero-state solution and the parametric dependence of the dynamics of the system.

\begin{figure}[th!]
  \begin{center}
  {\scalebox{\scl}{\includegraphics{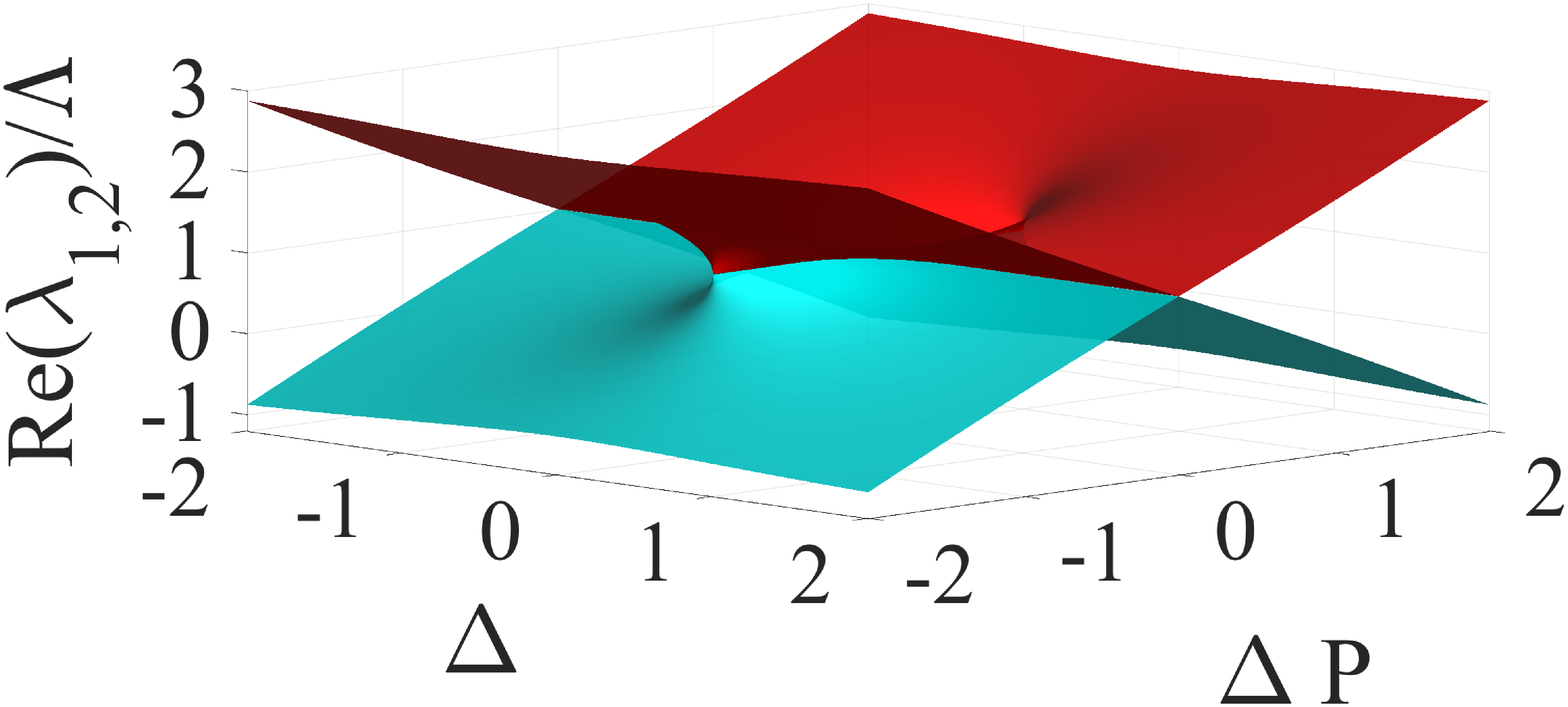}}} \hspace{-1em}
  {\scalebox{\scl}{\includegraphics{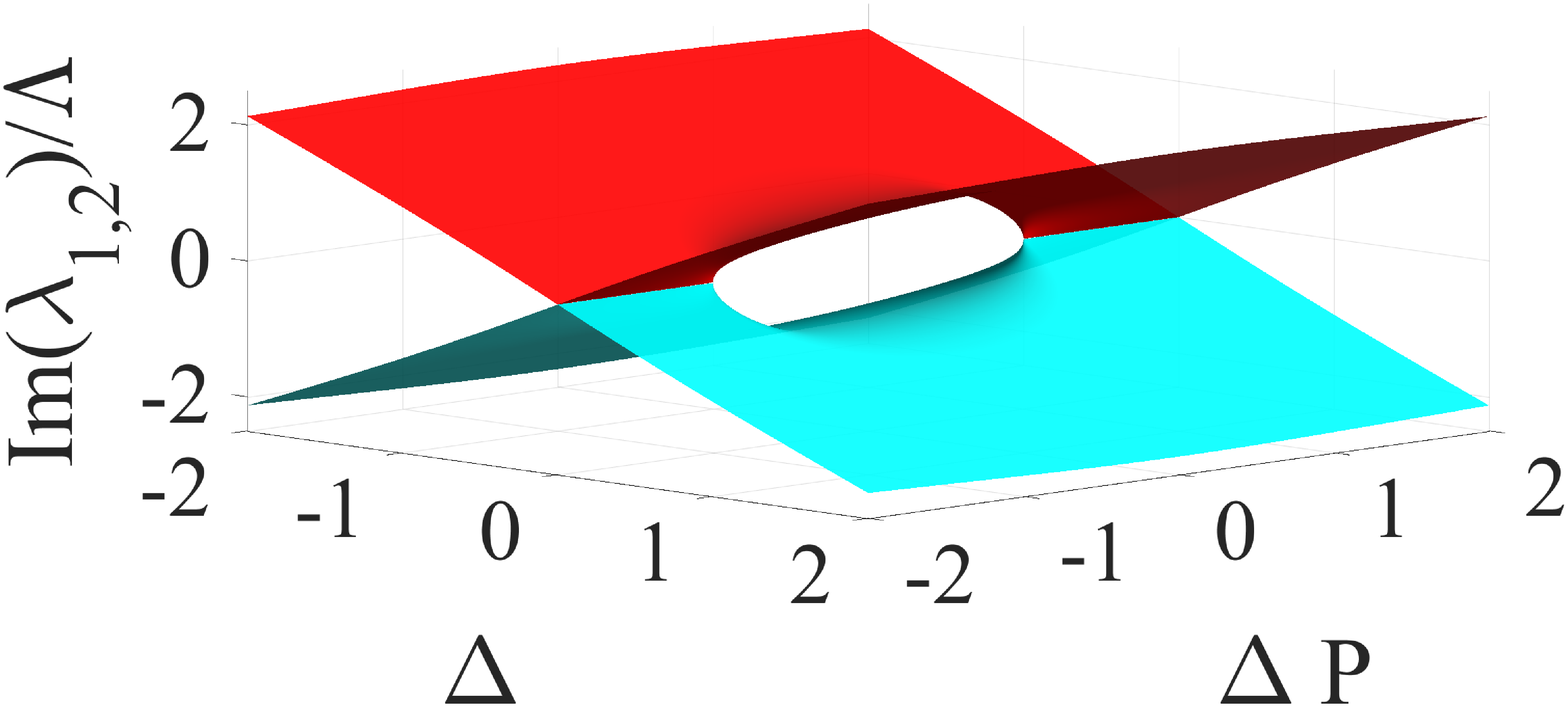}}}\\
  {\scalebox{\scl}{\includegraphics{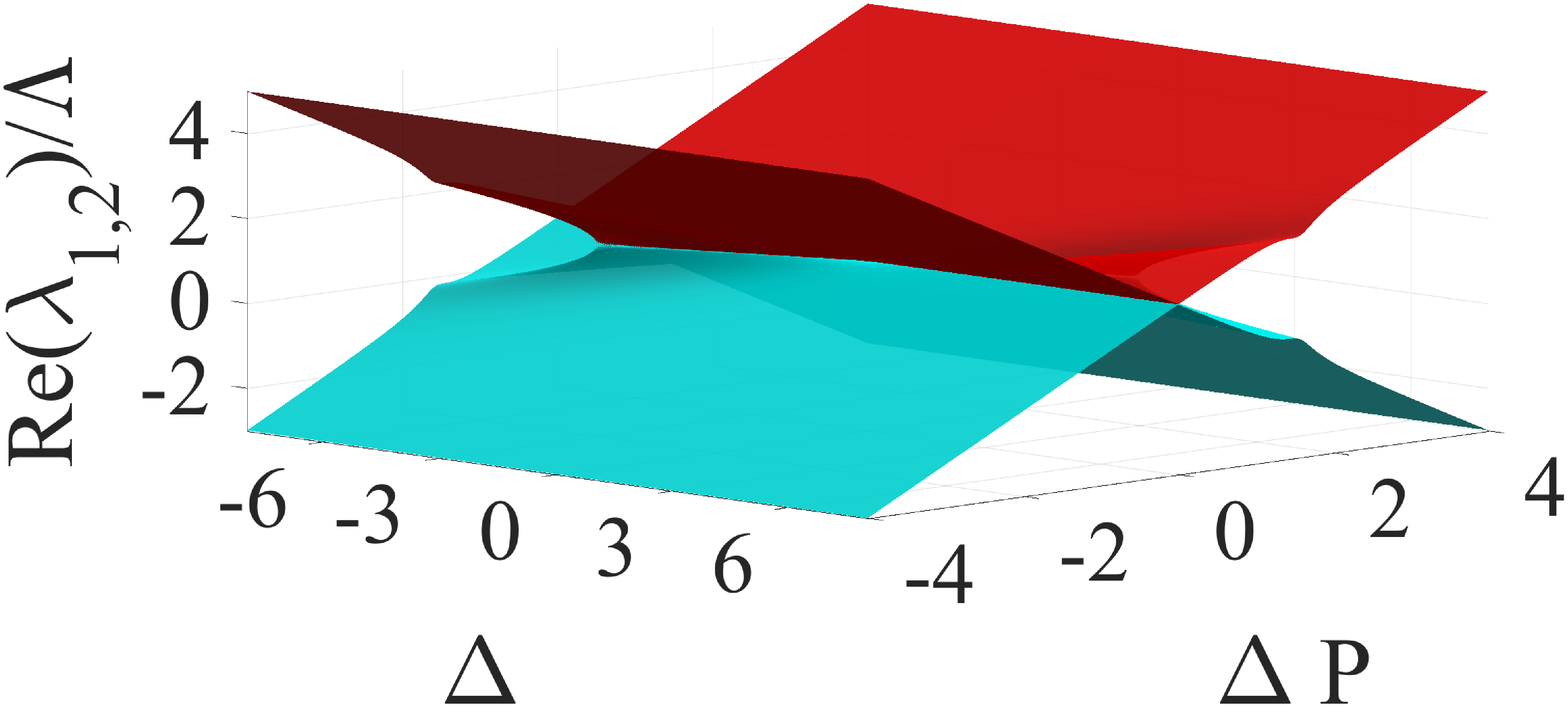}}}\hspace{-1em}
 {\scalebox{\scl}{\includegraphics{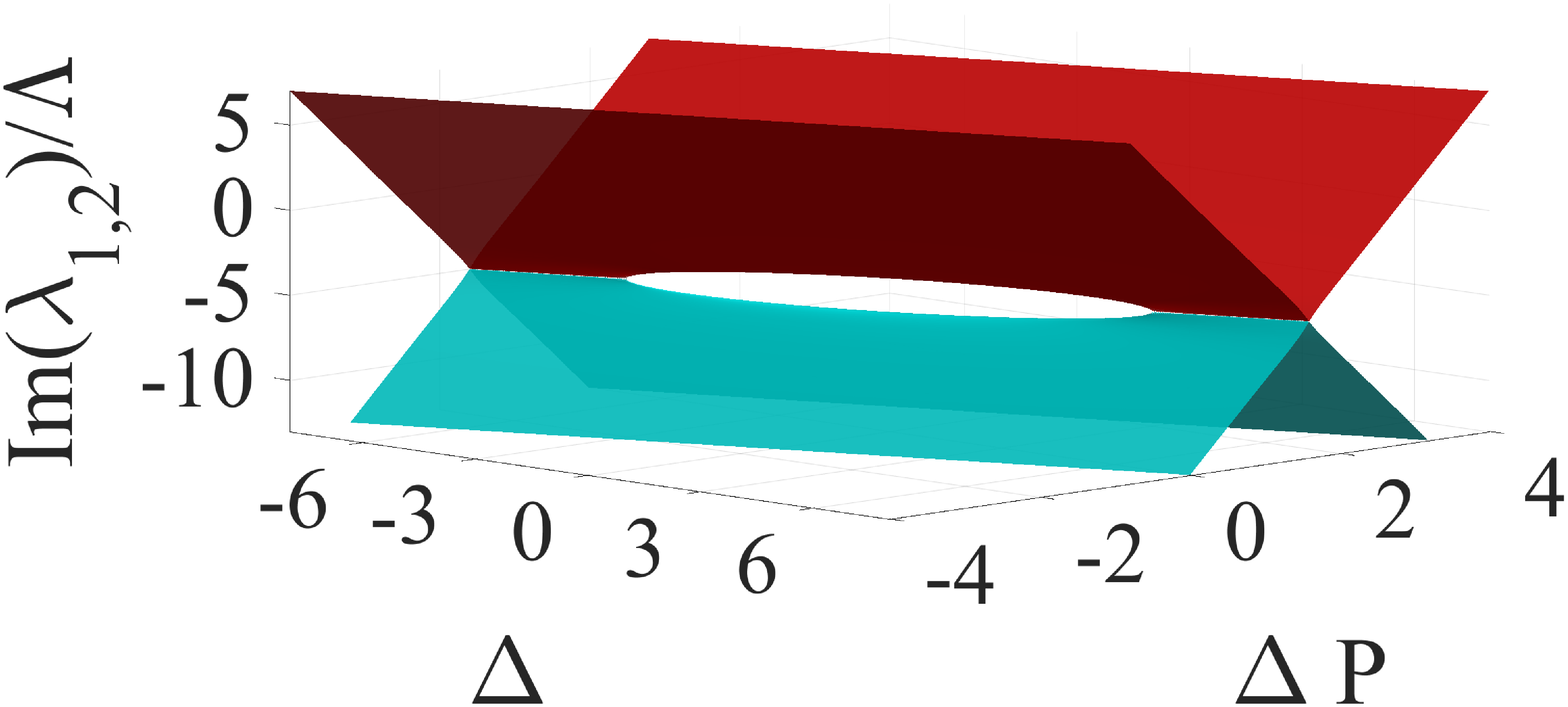}}}\\
  \caption{Real (left) and imaginary (right) part of the normalized eigenvalues $\lambda_{1,2}/\Lambda$ of the zero state for a zero ( $\alpha=0$, top) and a non-zero ($\alpha=5$ , bottom) linewidth enhancement factor. Spectral transitions and exceptional points existence take place along the line $\Delta-\alpha \Delta P=0$. A non-zero linewidth enhancement factor raises the restriction of zero detuning ($\Delta=0$), for the existence of exceptional points. }
  \end{center}
\end{figure}

The important role of the linewidth enhancement factor ($\alpha$) is obvious: a non-zero $\alpha$ suggests that both the pumping and the detuning determine the real and the imaginary part of the spectrum. This is intuitively expected, since the linewidth enhancement factor $(\alpha)$ introduces a dynamical coupling between the amplitude and the phase of the electric field. Moreover, the crucial role of the carrier density dynamics with respect to the spectral properties of the system is evident. It is worth emphasizing that these effects are absent in non-Hermitian dimers modelled by coupled-mode equations that neglect the nonlinearity of the system due to the coupling of the electric fields with the respective carrier densities. The real and imaginary parts of the eigenvalues $\lambda_{1,2}$ (normalized over the coupling constant $\Lambda$) as functions of the detuning $\Delta$ and pumping difference $\Delta P$ are depicted in Fig. 1. It is clear that spectral transitions, where real and imaginary parts of the eigenvalues coalasce, take place along the straight line $\Delta-\alpha \Delta P=0$ at the exceptional points, located at $\Delta P=\pm1$. This is a direct generalization of the case commonly considered in $PT$-symmetric non-Hermitian dimers, for which spectral transitions occur only along the zero detuning line $\Delta = 0$ \cite{Choquette_2017}. For the case of zero detuning $\Delta=0$ and for $\alpha=0$, spectral transitions take place analogously to the $PT$-symmetric case. For $|\Delta P|<1$, the eigenvalues $\lambda_{1,2}$ have a common nonzero real part and opposite imaginary parts along the line $\Delta=0$ whereas for $|\Delta P|>1$, the eigenvalues have different real parts and zero imaginary parts [Figs. 1(top)]. For a nonzero $\alpha=5$ there are no spectral transitions along $\Delta=0$. However, spectral transitions occur along the line $\Delta=\alpha \Delta P$ as shown in Fig. 1(bottom). One essential difference with the case of a zero $\alpha$ is that the imaginary parts of $\lambda_{1,2}$ are now symmetric, not with respect to zero, but with respect to a non-zero value. Therefore, for the realistic case of a nonzero $\alpha$, the asymmetric pumping can conctrol the spectral properties of the system.  \

\begin{figure}[pt]
  \begin{center}
  {\scalebox{\scl}{\includegraphics{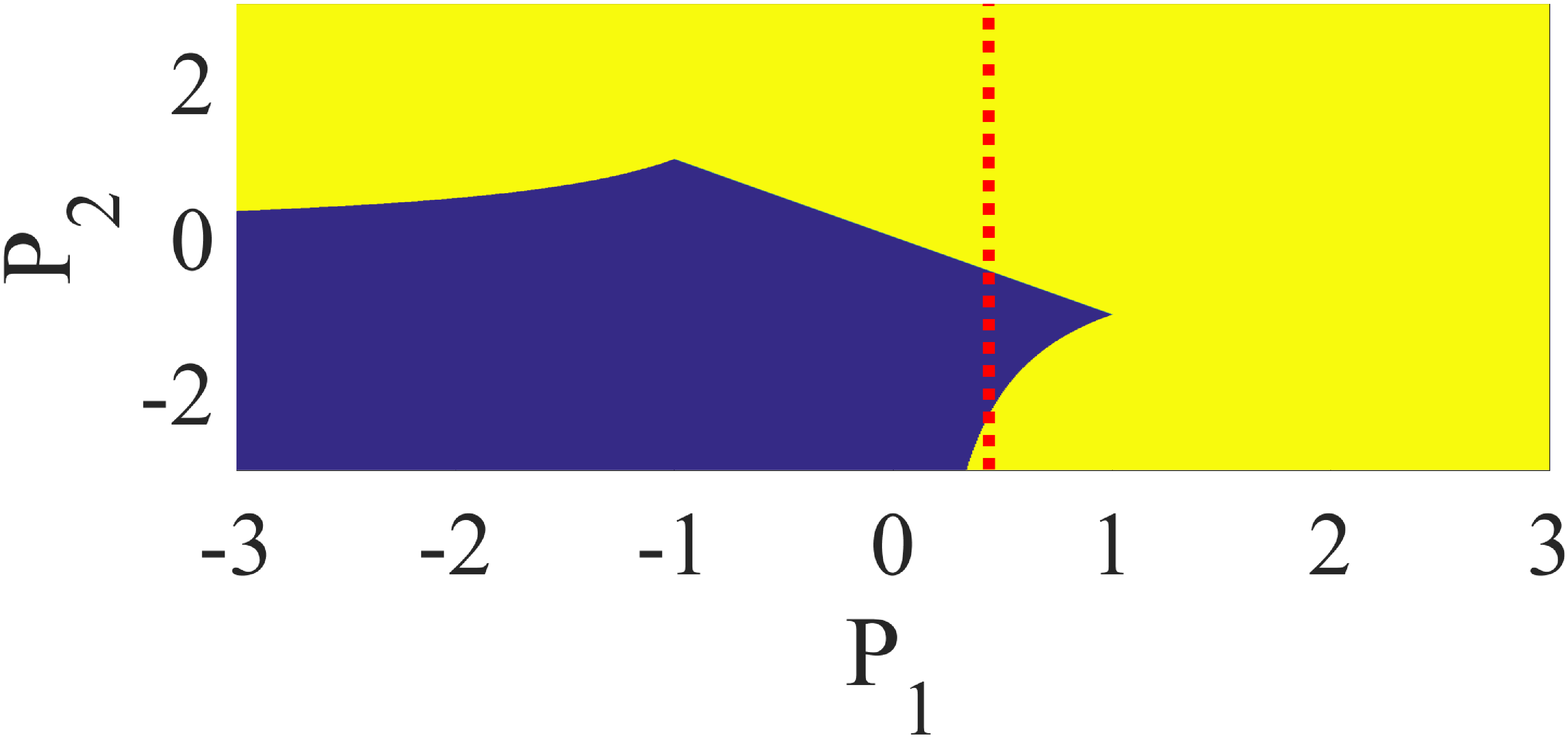}}} \hspace{-1em}
  {\scalebox{\scl}{\includegraphics{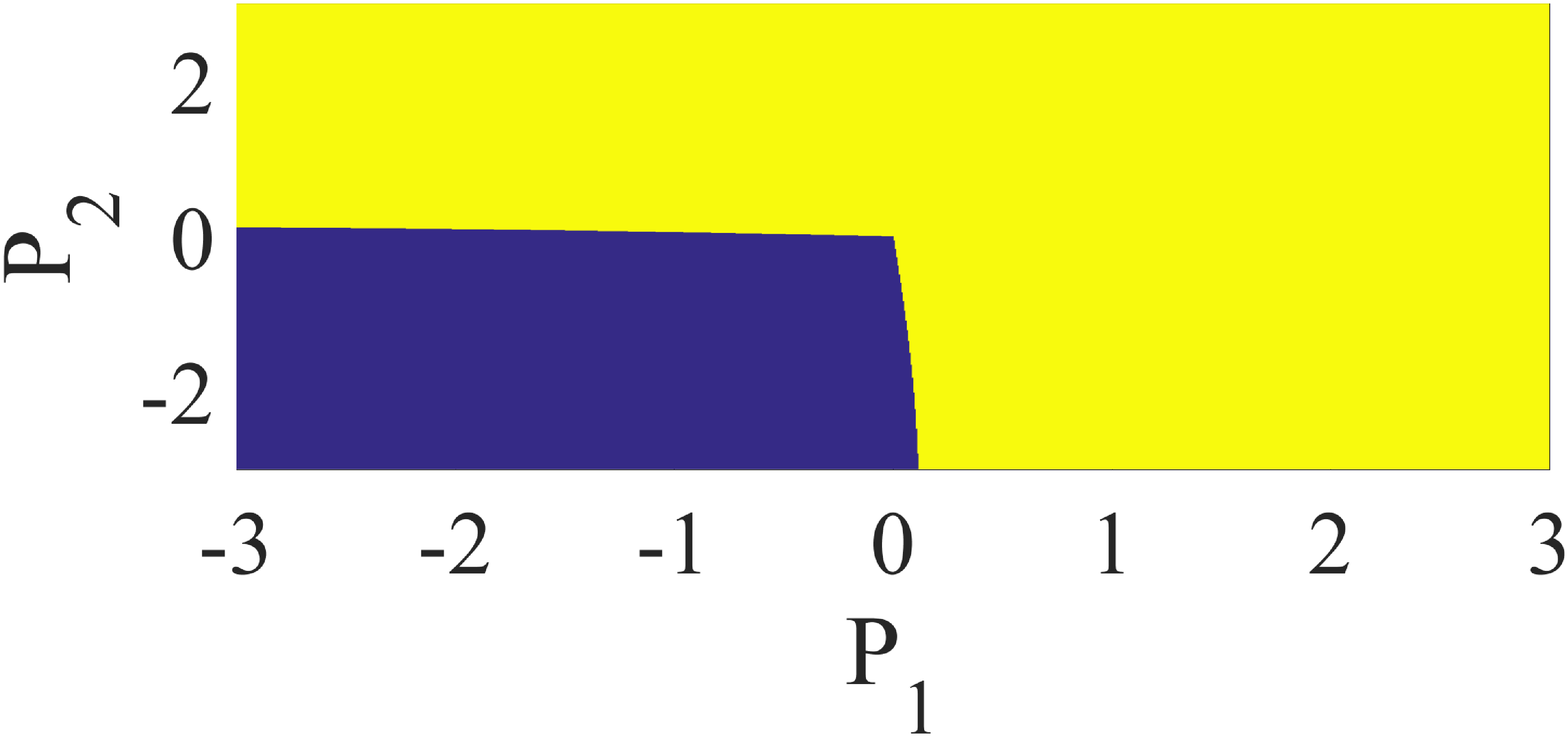}}}\\
  {\scalebox{\scl}{\includegraphics{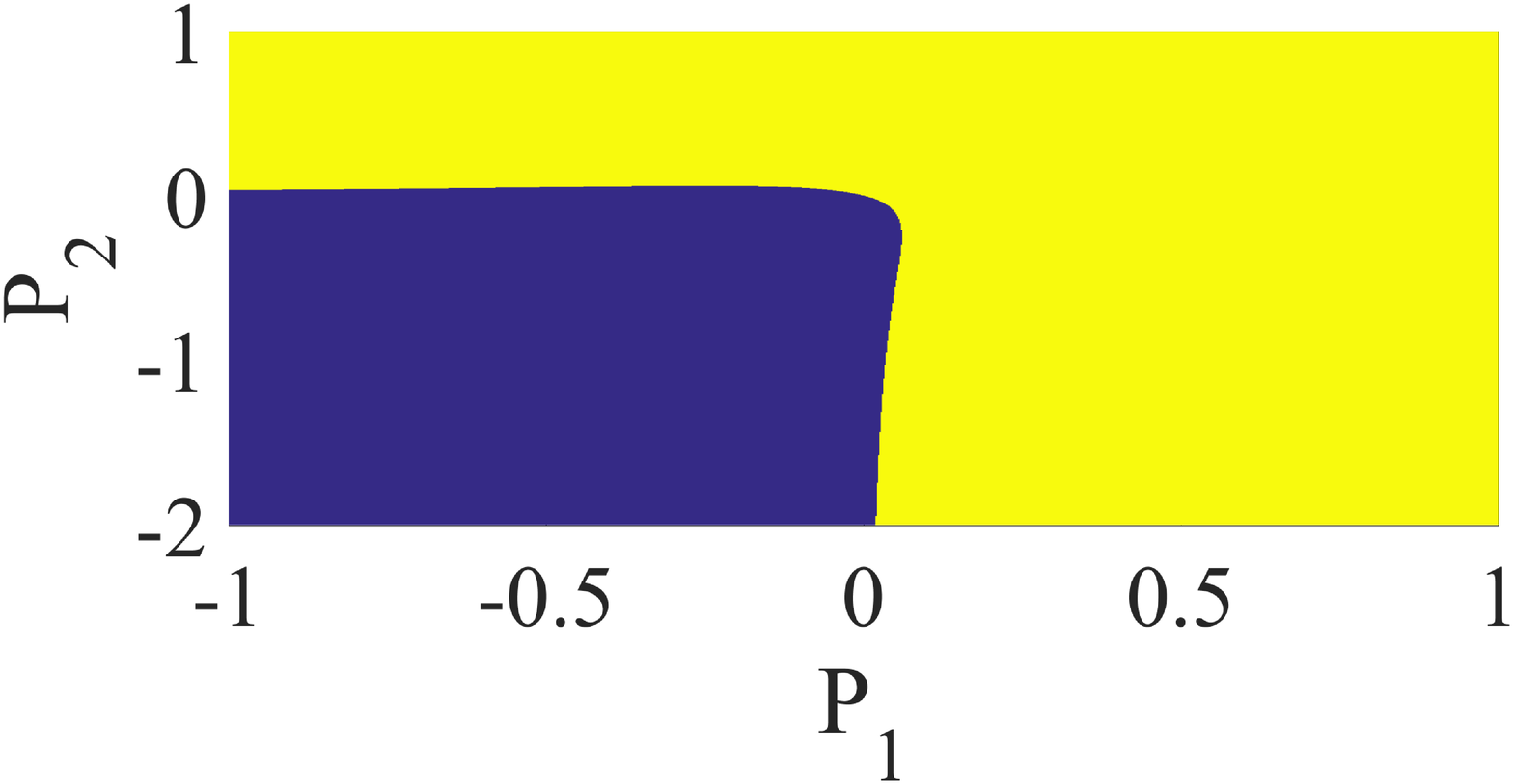}}} \hspace{-1em}
  {\scalebox{\scl}{\includegraphics{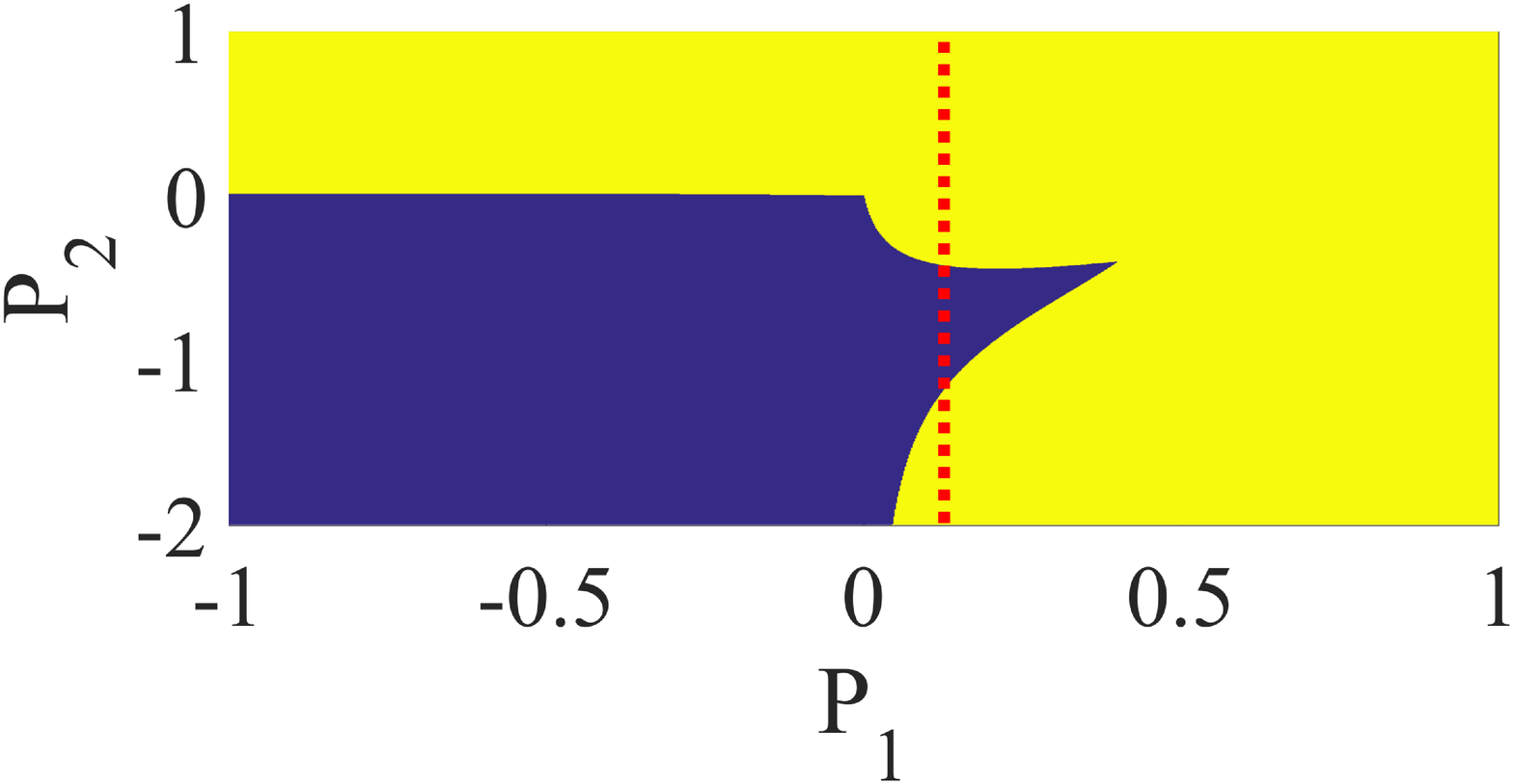}}}
   \caption{Stability region of the zero state in the ($P_1, P_2$) parameter space for zero [$\alpha=0$ (top)], and non-zero [$\alpha=5$ (bottom)] linewidth enhancement factor, under zero [$\Delta=0$ (left)] and non-zero [$\Delta=2$ (right)] detuning. The red dotted lines correspond to $P_1=0.5$ (a) and $P_1=0.15$ (d). Reversing of laser pump dependence and self-termination scenarios take place for increasing $P_2$ along the constant $P_1$ lines.}
  \end{center}
\end{figure}

\begin{figure}[h]
  \begin{center}
  {\scalebox{\scl}{\includegraphics{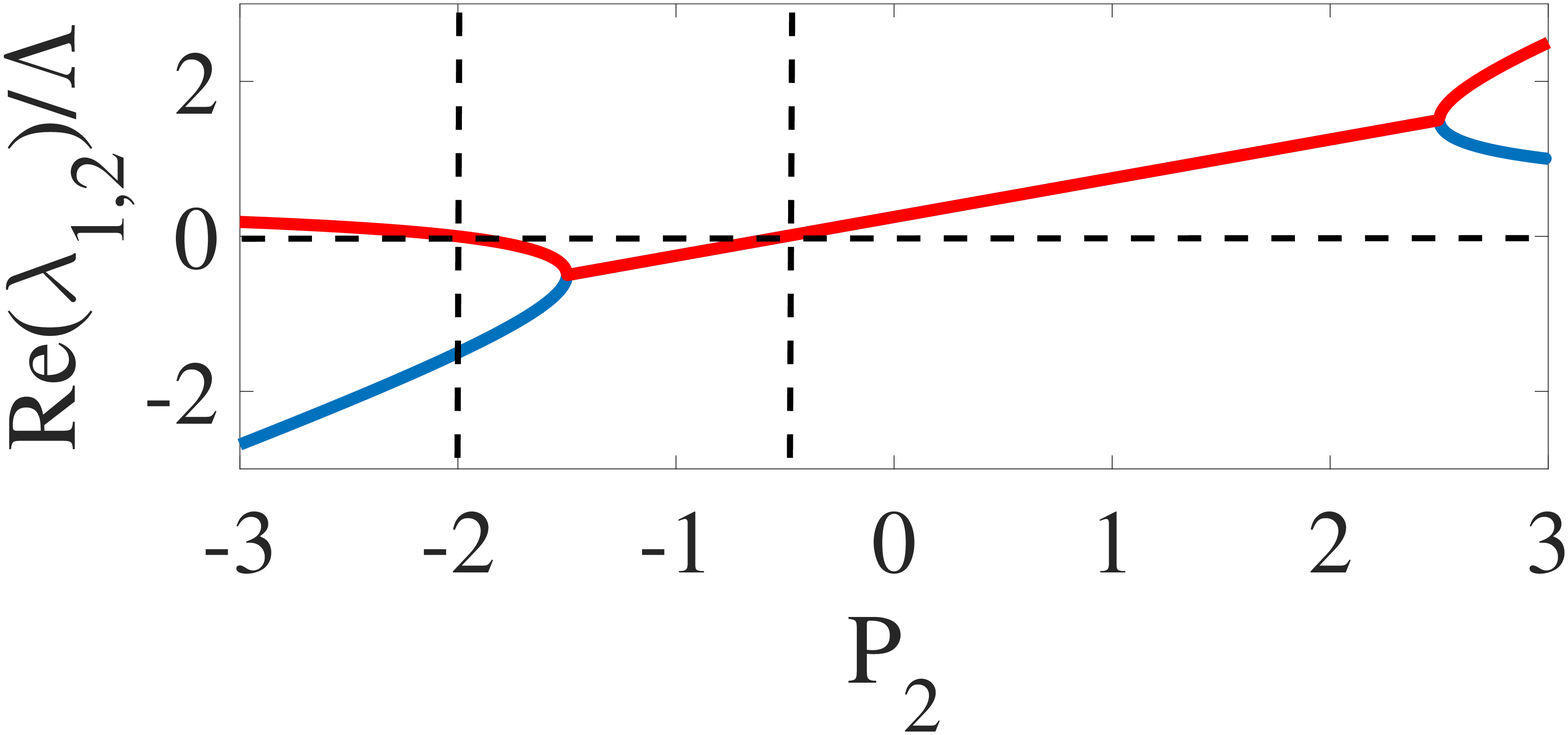}}} \hspace{-1em}
  {\scalebox{\scl}{\includegraphics{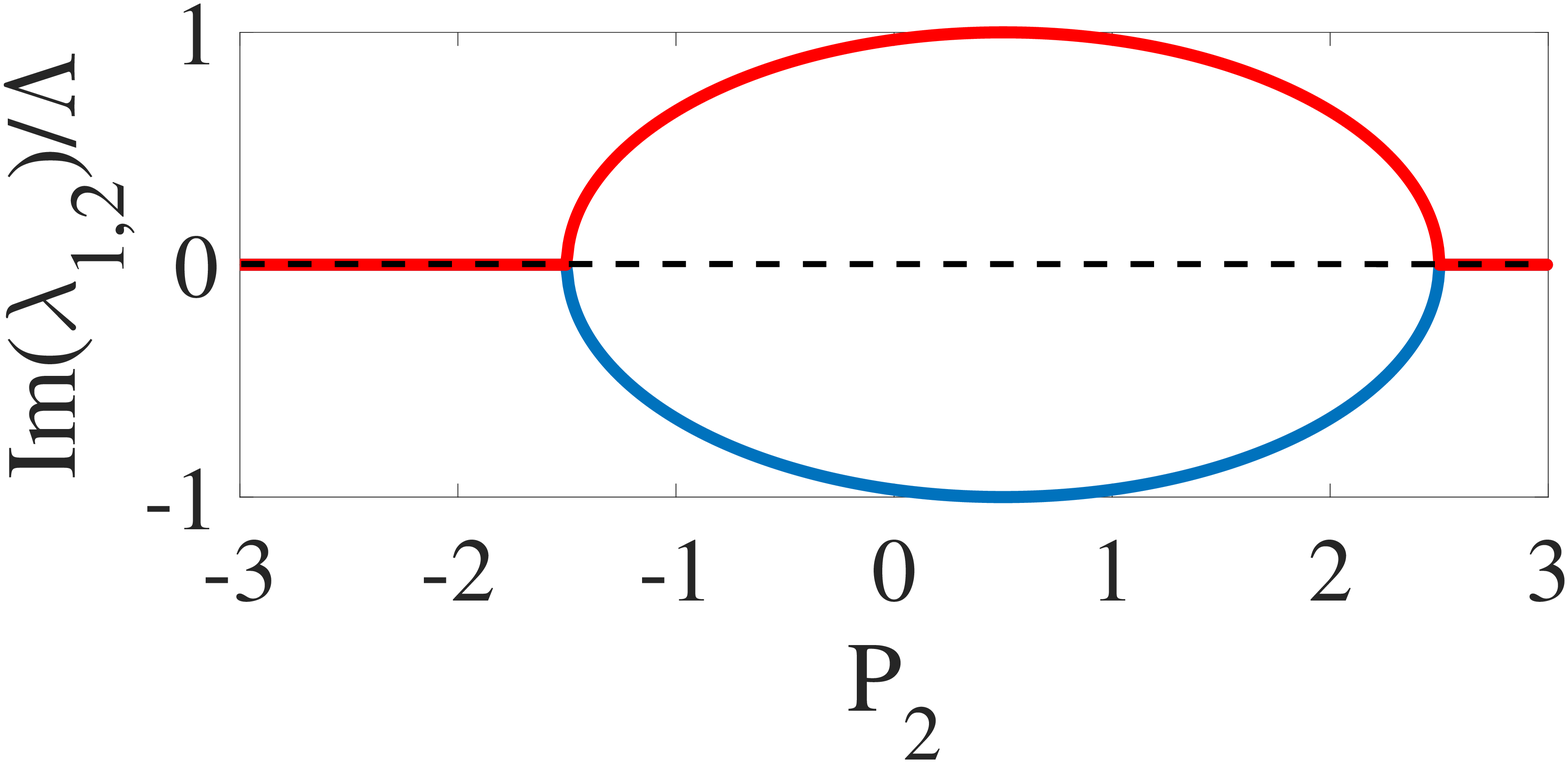}}}\\ 
  {\scalebox{\scl}{\includegraphics{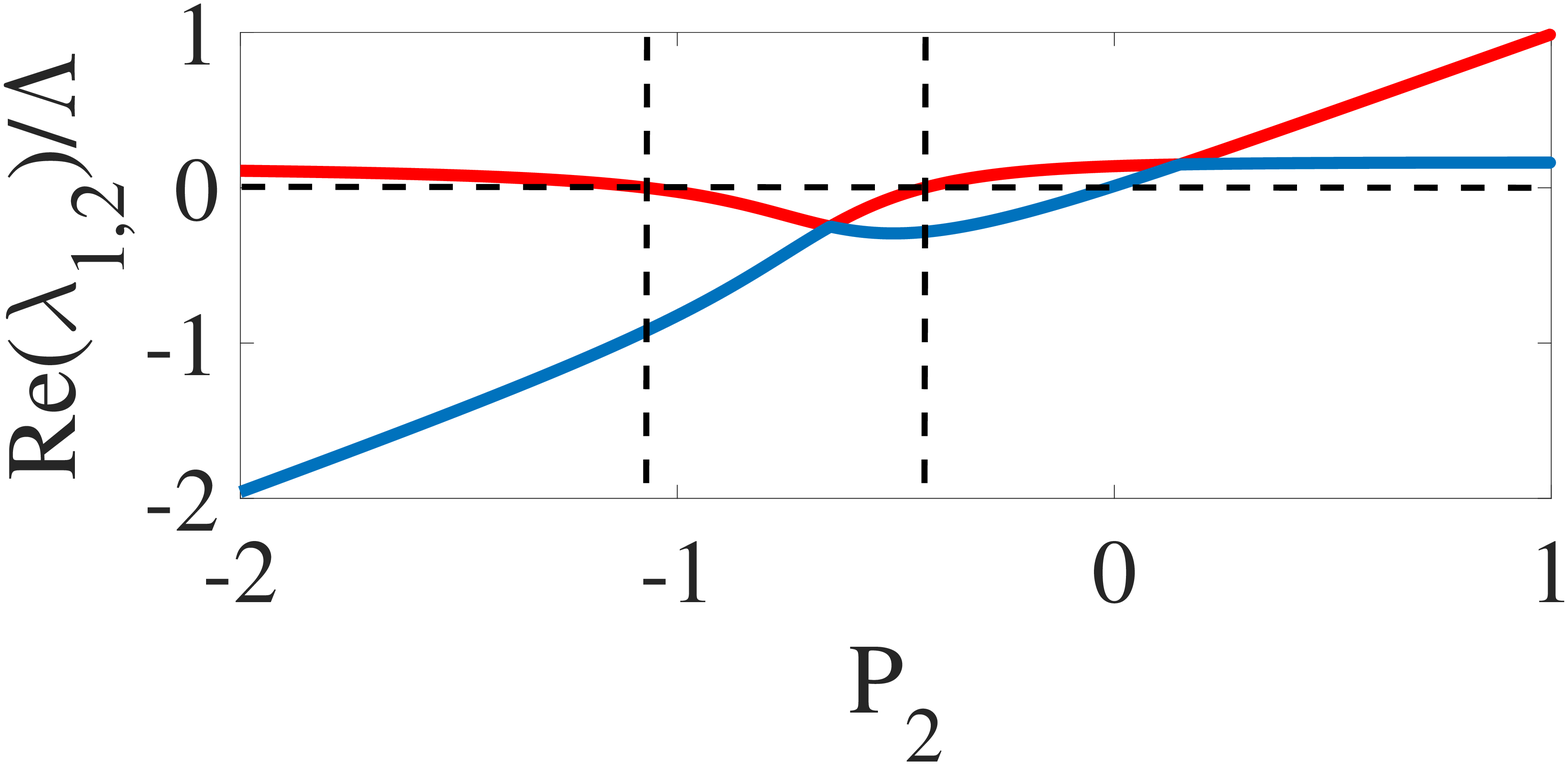}}} \hspace{-1em}
  {\scalebox{\scl}{\includegraphics{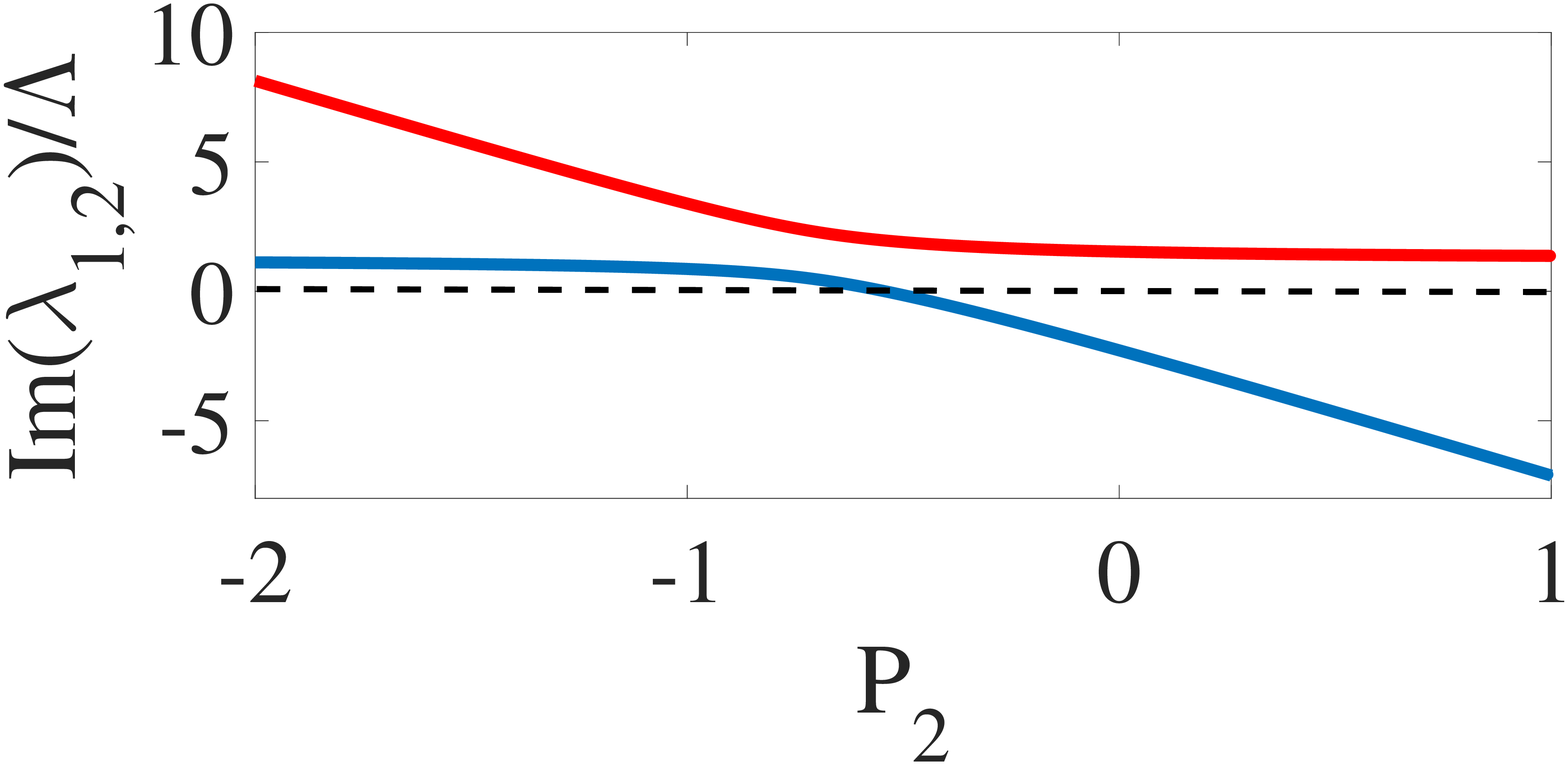}}}
   \caption{Eigenvalues trajectories corresponding to the cases of Fig. 2(top, left) and Fig. 2(bottom, right). The vertical dashed lines denote zero crossings of the real part and stability changes of the zero state. The rate equation model predicts that self-termination scenarios can take place under configurations not restricted by exact or approximate fulfilment of PT-symmetry conditions. }
  \end{center}
\end{figure}

The stability of the zero state determines the lasing conditions of the dimer. The zero state becomes unstable when one of the eigenvalues $\lambda_{1,2}$ has a positive real part. In such case, the system evolves either to a stable nonzero state, if such a state exists for the specific parameter set, or to a limit cycle, or to a chaotic state \cite{Kovanis_PRL14}. The dependence of the zero state stability  on the pumping rates $P_{1,2}$ for a zero ($\alpha=0$) and a non-zero ($\alpha=5$) linewidth enhancement factor is depicted in Fig. 2. The role of detuning is shown to crucially determine the shape of the stable region in the $(P_1,P_2)$ parameter space. For $\alpha=0$ the stability region is symmetric with respect to the line $P_1=P_2$ [Figs. 2(top)] and tends to a rectangular shape for an increasing detuning as shown in Fig. 2(top, right). For the case of non-zero linewidth enhancement factor ($\alpha=5$) the stability region has a rounded rectangular shape for zero detuning [Fig. 2(bottom, left)] and is asymmetrically deformed for a non-zero ($\Delta=2$) detuning as shown in Fig. 2(bottom, right) (for a $\Delta$ of opposite sign, a deformation that is symmetric with respect to the line $P_1=P_2$ is obtained). The shape of the stability regions dictates the bifurcation scenarios of the zero state for varying $P_1$ and $P_2$. For the case of $\alpha=0$, $\Delta=0$ and a fixed $P_1=0.5$ [red dotted line in Fig. 2(top, left)], an increasing $P_2$ results in stability changes for the zero state, so that we have lasing for small and large values of $P_2$ where the zero state is unstable, whereas there exist a range of $P_2$ values for which the zero state is stable and the lasing is terminated. The same scenario occurs for $\alpha=5$, $\Delta=2$ and $P_1=0.15$ [red dotted line in Fig. 2(bottom, right)]. Thus, it is the stability of the zero state and its bifurcations that explain the experimentally observed effect of the reversing of the pump dependence and the self-termination of a pair of coupled lasers \cite{Rotter_2012, Rotter_2014, Peng_2014, El-Ganainy_2014}. The situation is further explained in Fig. 3, where the respective trajectories of the eigenvalues for an increasing $P_2$ is shown. The lasing is terminated for $P_2$ values between the vertical dashed lines denoting a sign change of the real part, where the zero state is stable. It is clear that the self-termination scenario can take place if the imaginary part of the two eigenvalues either coalesce as in the first case or simply approach each other as in the second case. Therefore, the rate-equation model predicts that self-termination scenarios can take place under much more general configurations than those implied by conditions of exact or approximate PT-symmetry, determined by simplified coupled mode equation models \cite{PT_1, PT_2, PT_3, PT_4, Ramezani_2010}, since a non-zero linewidth enhancement factor suggests that spectral transitions are not restricted to the zero detuning line $\Delta=0$. \


For the study of the nonzero states of the system we introduce the amplitude and phase of the complex electric field amplitude in each laser as $\mathcal{E}_i=X_ie^{i\theta_i}$.
For a given phase-locked state with field amplitude ratio $\rho \equiv X_2/X_1$ and phase difference $(\theta)$ we can solve the algebraic system of equations obtained by setting the rhs of Eq. (\ref{array}) equal to zero, and obtain the steady-state carrier densities $(Z_{1,2})$, and the appropriate detuning $(\Delta)$ and pumping rates $(P_{1,2})$, in terms of $\rho$ and $\theta$ as follows:
\begin{eqnarray}
Z_1&=&\Lambda \rho \sin\theta, \hspace{2em} Z_2=-(\Lambda/\rho)\sin\theta \label{Z_eq}\\
\Delta&=&-\alpha \Lambda\sin\theta\left(1/\rho+\rho\right)-\Lambda\cos\theta\left(1 / \rho-\rho\right)  \label{D_eq}\\
 P_1&=&X_0^2+(1+2X_0^2)\Omega\Lambda\rho\sin\theta  \nonumber\\
 P_2&=&\rho^2X_0^2-(1+2\rho^2 X_0^2) (\Omega\Lambda/\rho)\sin\theta  \label{P_eq}
\end{eqnarray}
where $\Delta=\Omega_2-\Omega_1$ is the detuning, $\theta=\theta_2-\theta_1$ is the phase difference of the electric fields and we have used a reference value $P=(P_1+P_2)/2$ in order to define $\Omega$. 
Note that in the absence of detuning $(\Delta=0)$, for every $\rho$ the phase-locked state has a fixed phase-difference given by $\tan\theta= \frac{1}{\alpha} \frac{\rho^2-1}{\rho^2+1}$ but it has an arbitrary power $X_0$. If, in addition to zero detuning, we consider equal pumping $(P_1=P_2)$, then the power is fixed to the value $X_0^2 = \frac{\Omega \Lambda \sin\theta (\rho^2+1)}{\rho\left[(\rho^2-1)-4\Omega \Lambda \rho \sin\theta\right]} $. For nonzero detuning $(\Delta\neq0)$ and unequal pumping $(P_1\neq P_2)$, there exist phase-locked states with arbitrary amplitude asymmetry $(\rho)$, phase difference $(\theta)$ and power $(X_0)$ \cite{PRA_paper}.\ 

\begin{figure}[t!]
  \begin{center}
   {\scalebox{\scl}{\includegraphics{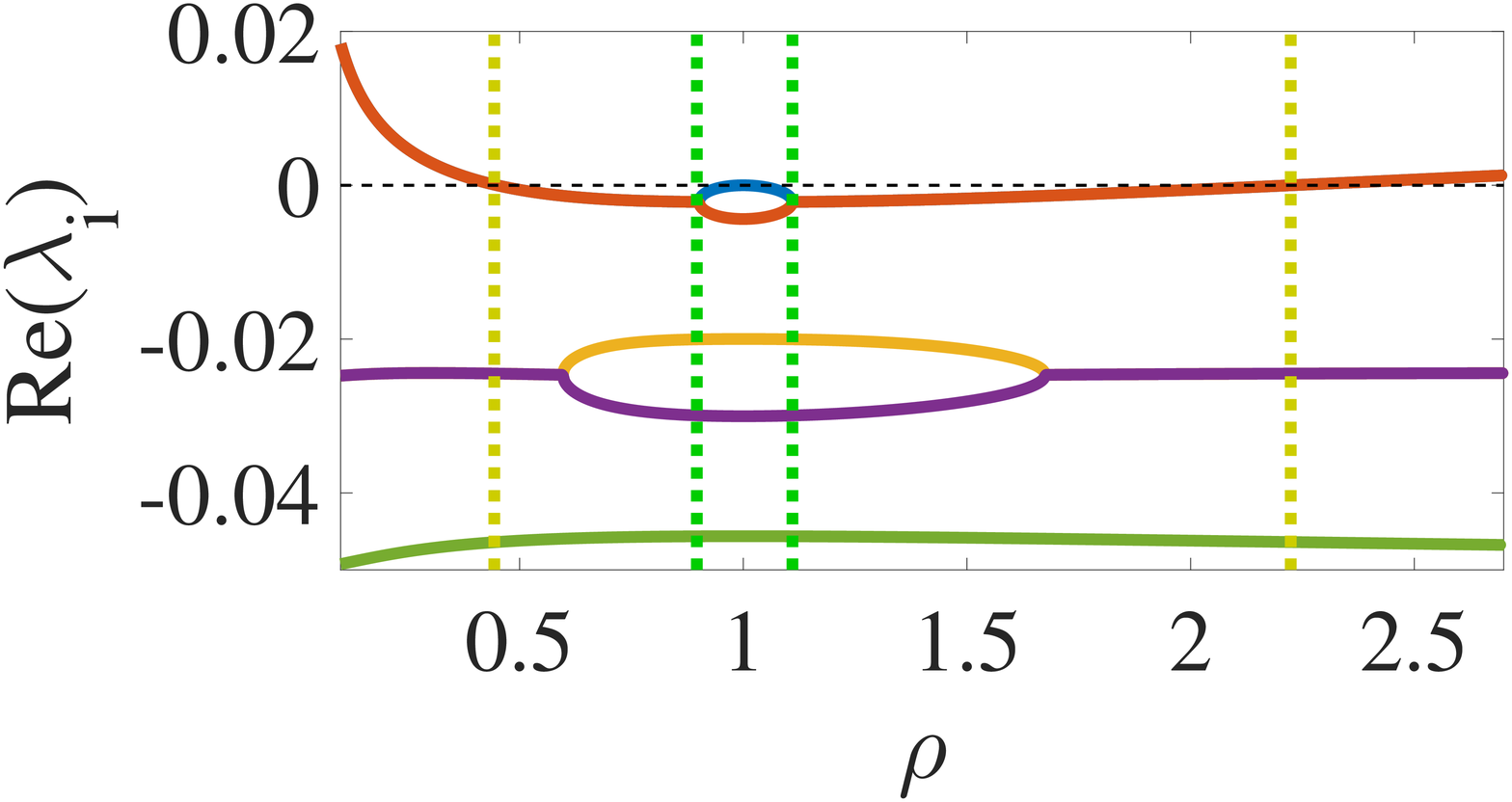}}} \hspace{-1em}
  {\scalebox{\scl}{\includegraphics{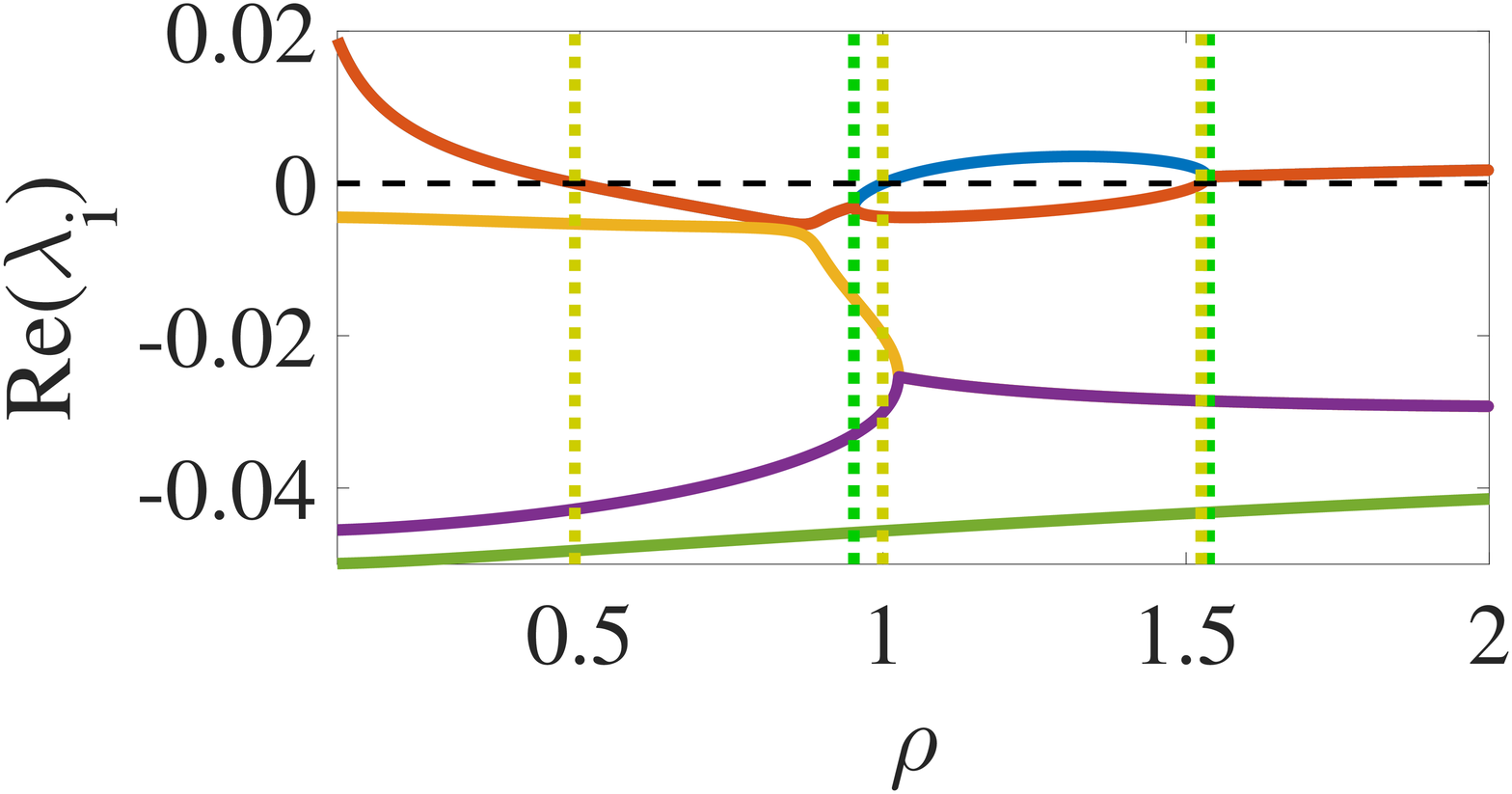}}}\\
    {\scalebox{\scl}{\includegraphics{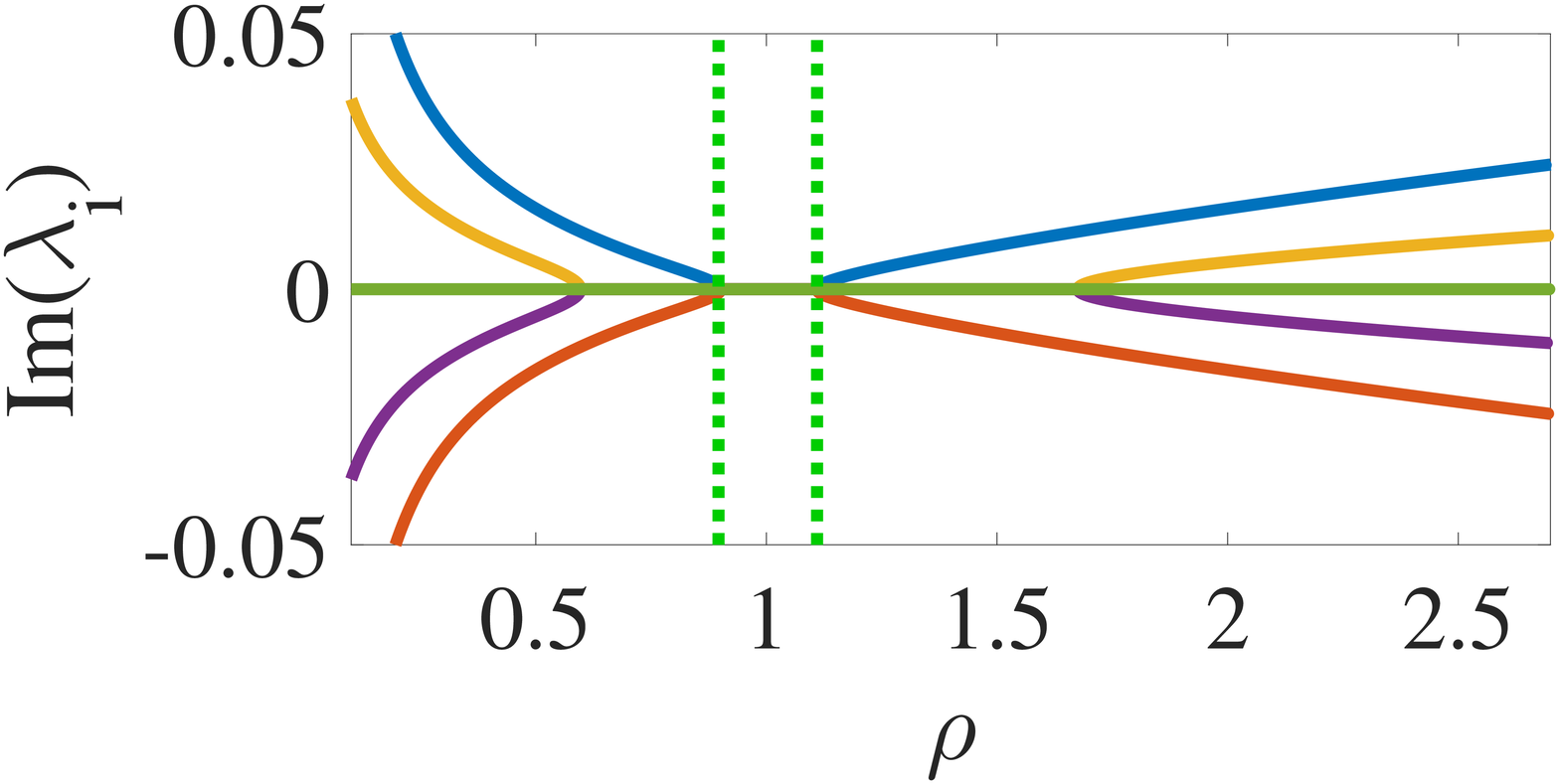}}} \hspace{-1em}
  {\scalebox{\scl}{\includegraphics{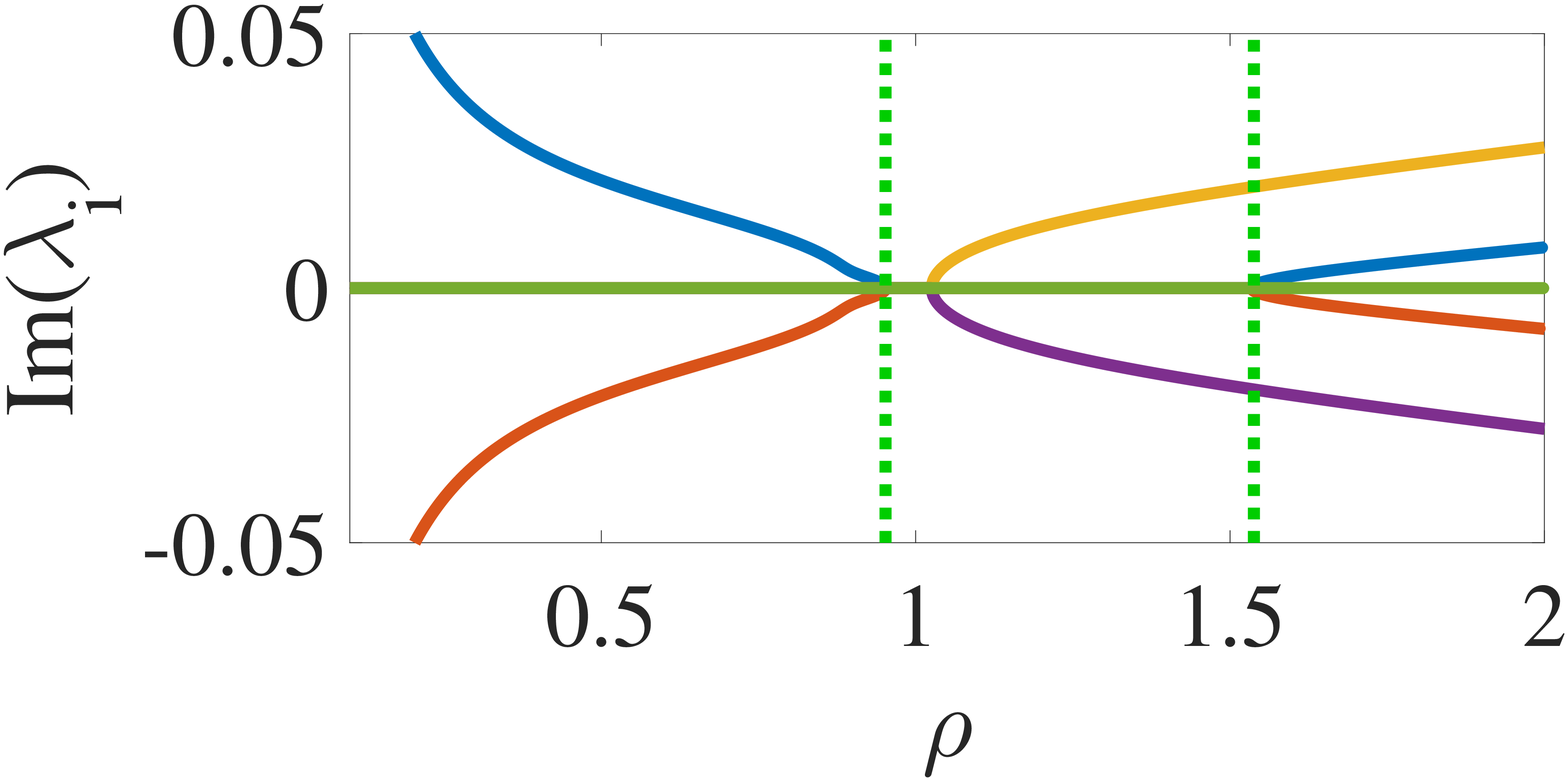}}}\\
      {\scalebox{\scl}{\includegraphics{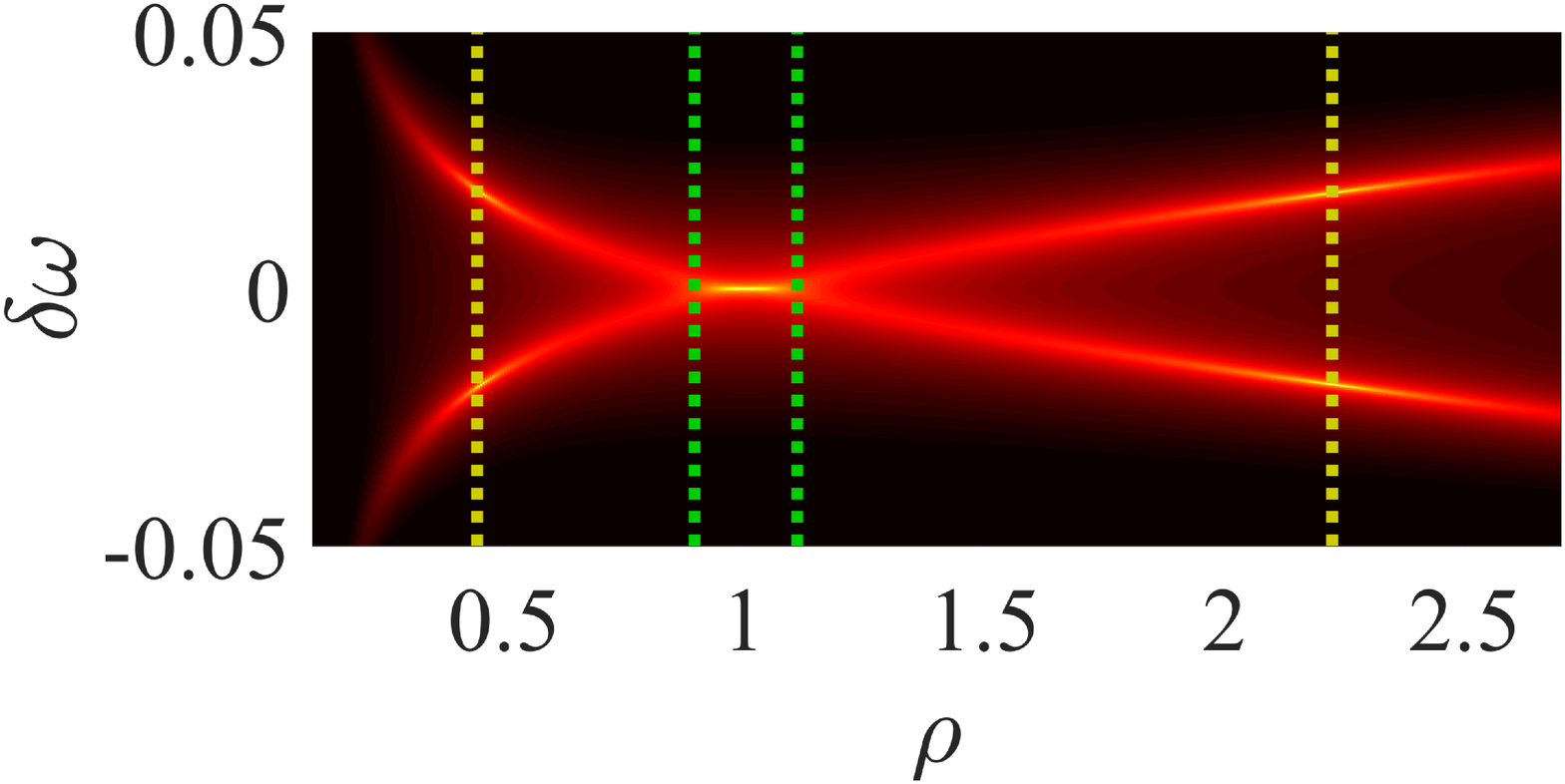}}} \hspace{-1em}
  {\scalebox{\scl}{\includegraphics{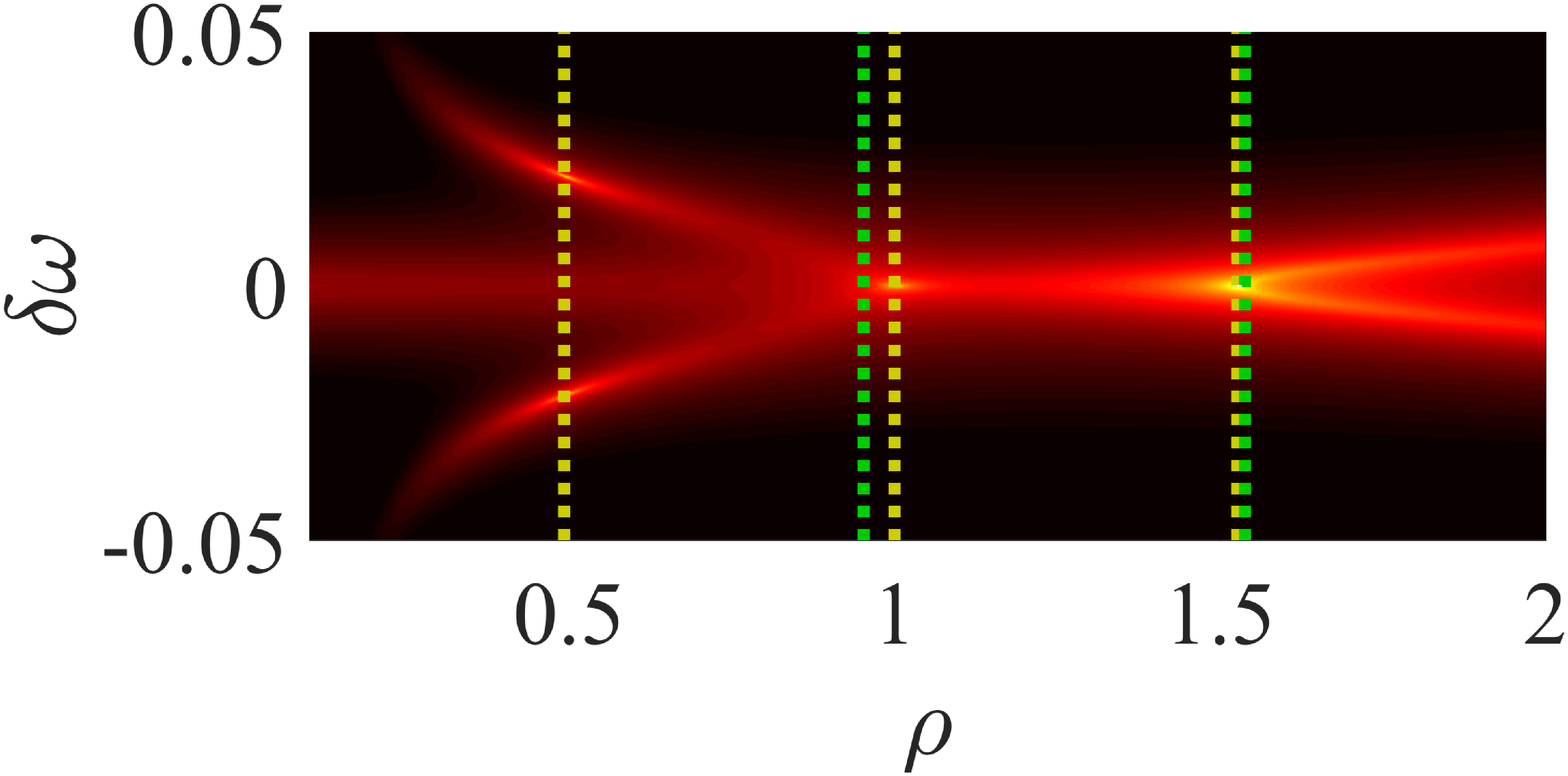}}}\\
   \caption{Eigenvalues (top, middle) and spectral line shape (bottom) of the phase-locked states for $\alpha=5$, $\Delta=0$ and $\log\Lambda=-2$. The vertical yellow and green dotted lines correspond to zero-crossings of the eigenvalue real part and exceptional points, respectively. (left) Phase-locked states under symmetric pumping $P_1=P_2$ ($\theta$ and $X_0$ are determined by $\rho$). The phase-locked states are stable between the vertical yellow dotted lines. (right) Phase-locked states under asymmetric pumping $P_1\neq P_2$ ($\theta$ is determined by $\rho$ and $X_0=0.01$). The phase-locked states are stable between the first and the second (left to right) vertical yellow dotted lines. The spectral line shape follows the imaginary part of the eigenvalues. Exceptional and bifurcation points are manifested in the spectral line shape through the emergence of side bands and intensity peaks, respectively.}
  \end{center}
\end{figure}

\begin{figure}[t!]
  \begin{center}
  {\scalebox{\scl}{\includegraphics{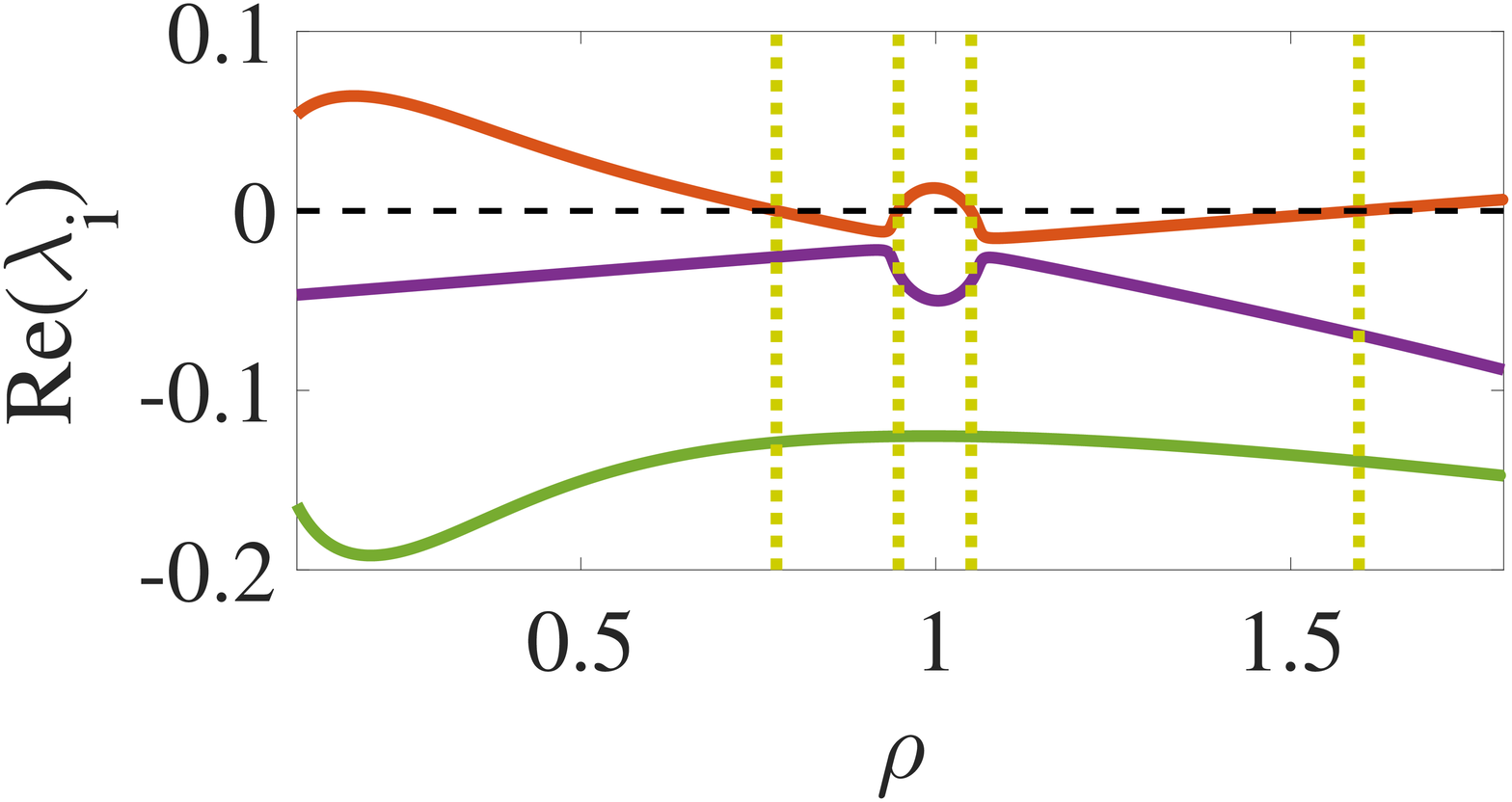}}} \hspace{-1em}
  {\scalebox{\scl}{\includegraphics{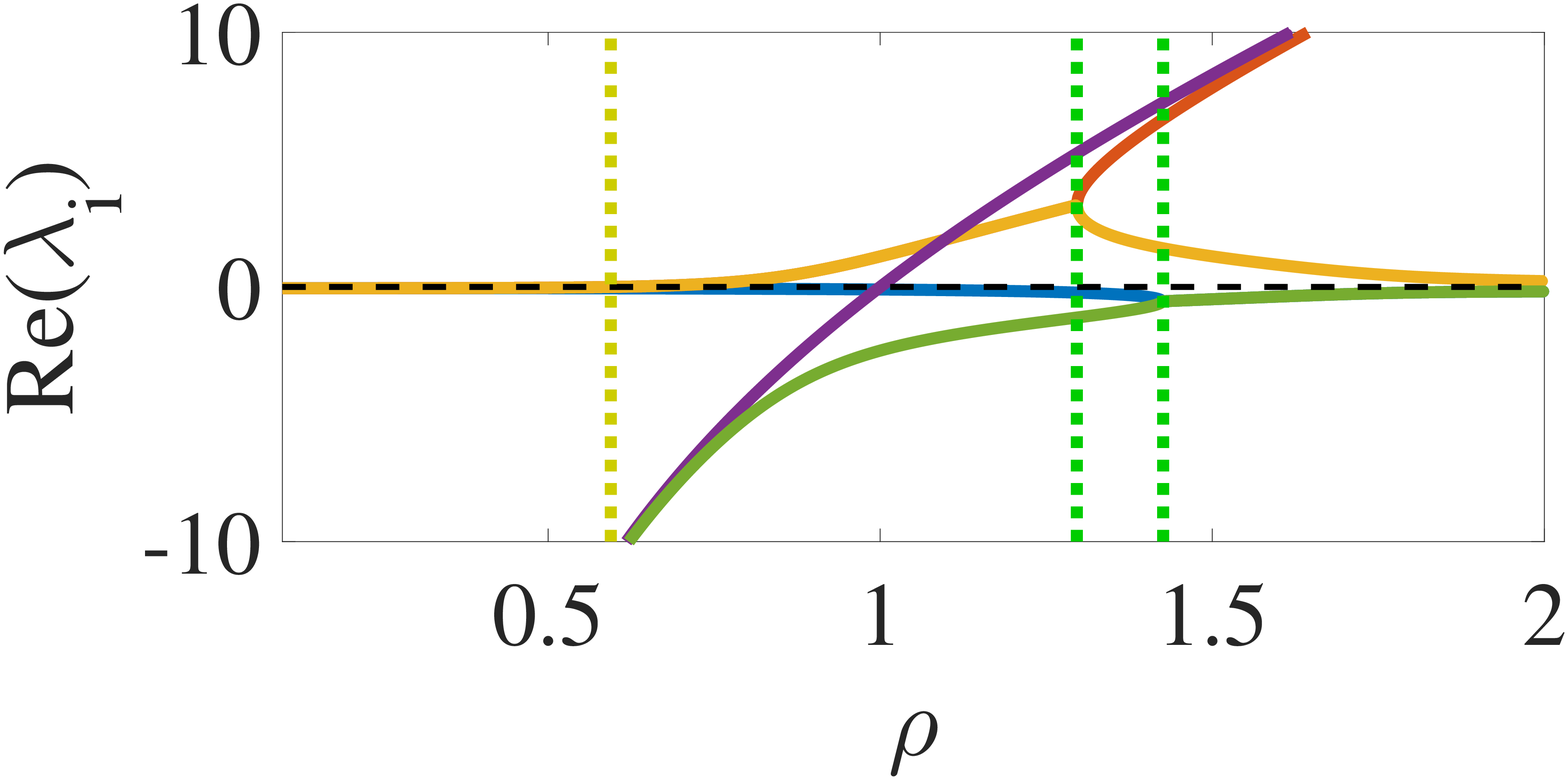}}}\\
  {\scalebox{\scl}{\includegraphics{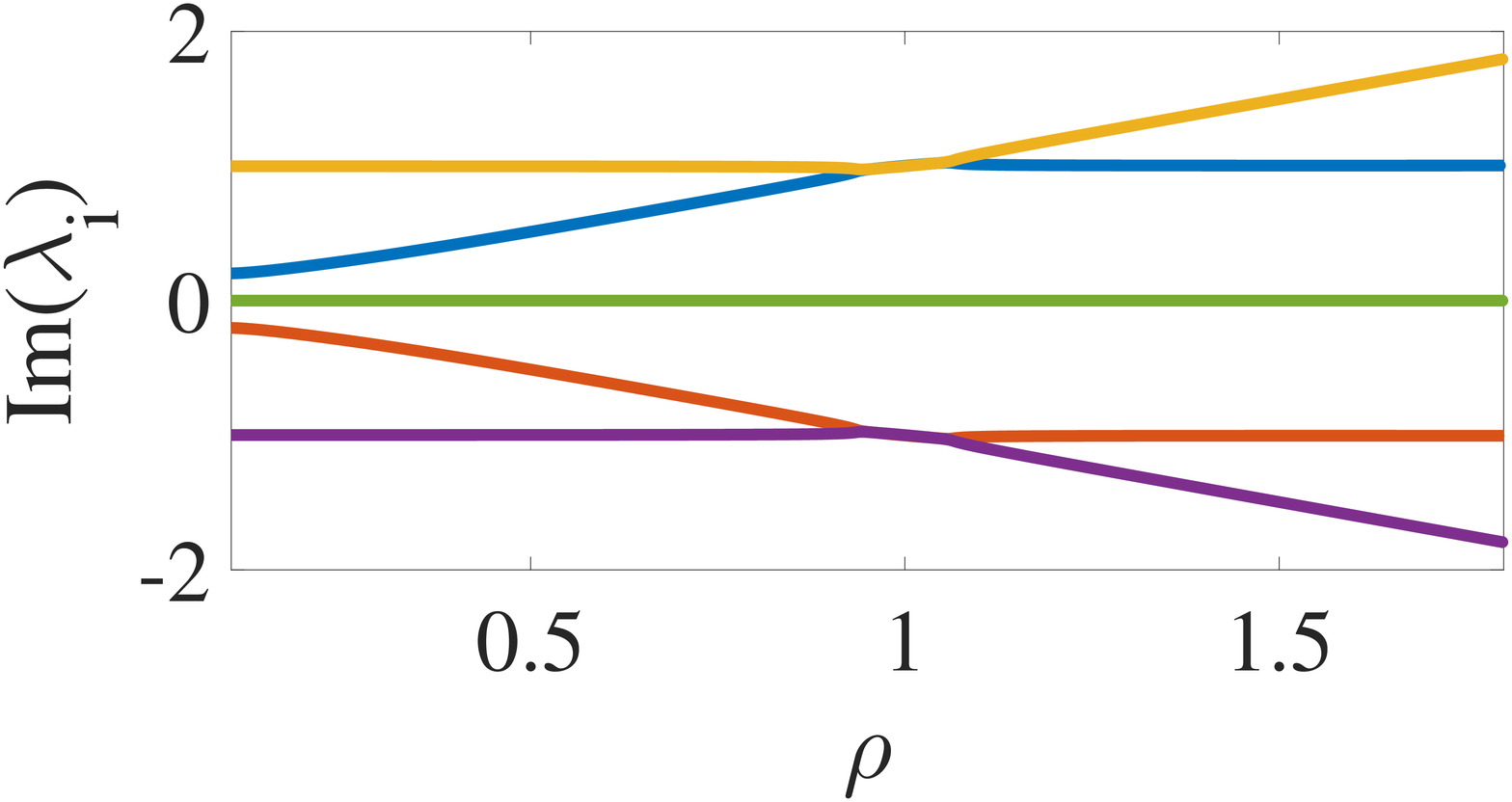}}} \hspace{-1em}
  {\scalebox{\scl}{\includegraphics{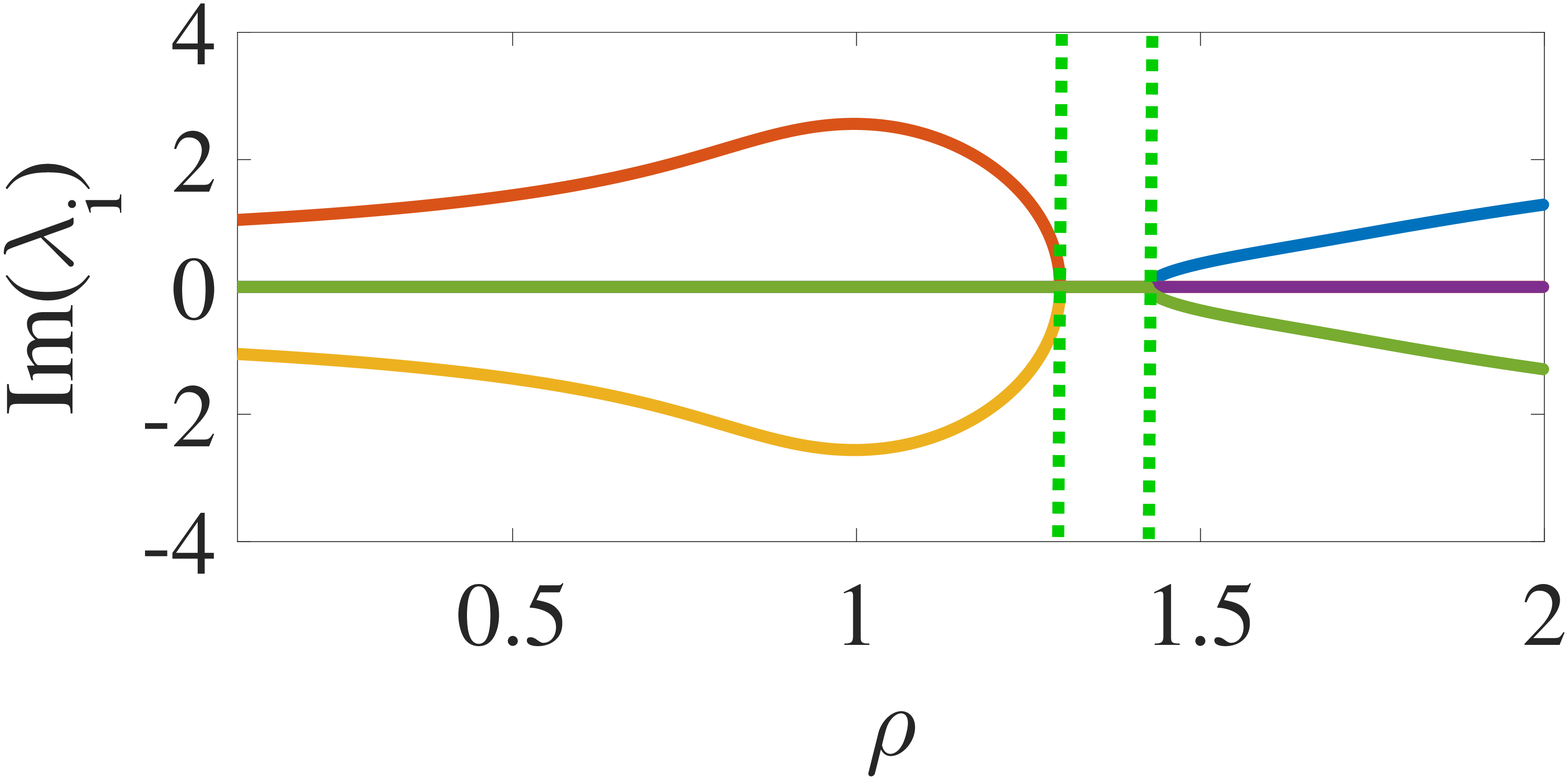}}}\\
   {\scalebox{\scl}{\includegraphics{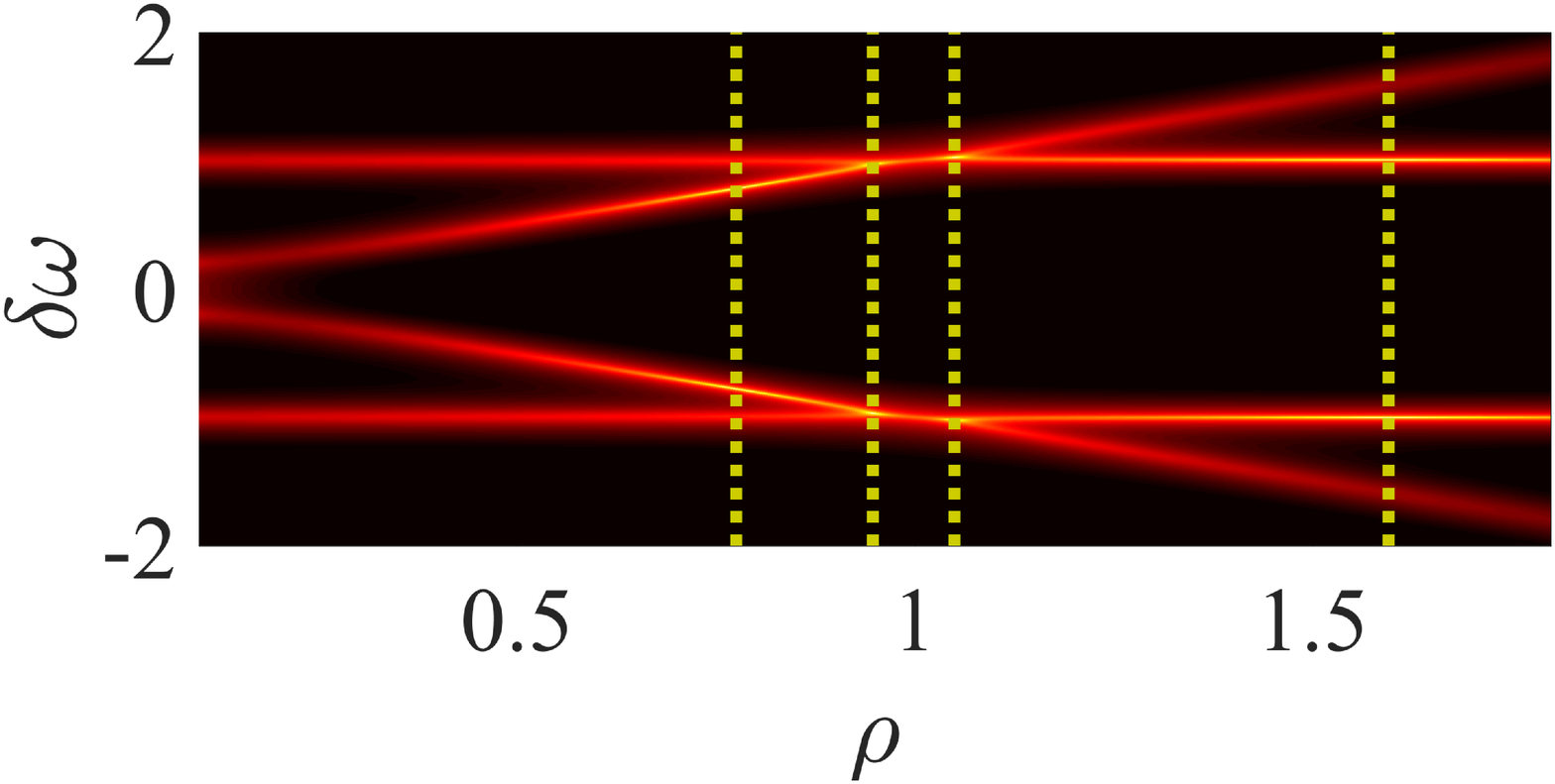}}} \hspace{-1em}
  {\scalebox{\scl}{\includegraphics{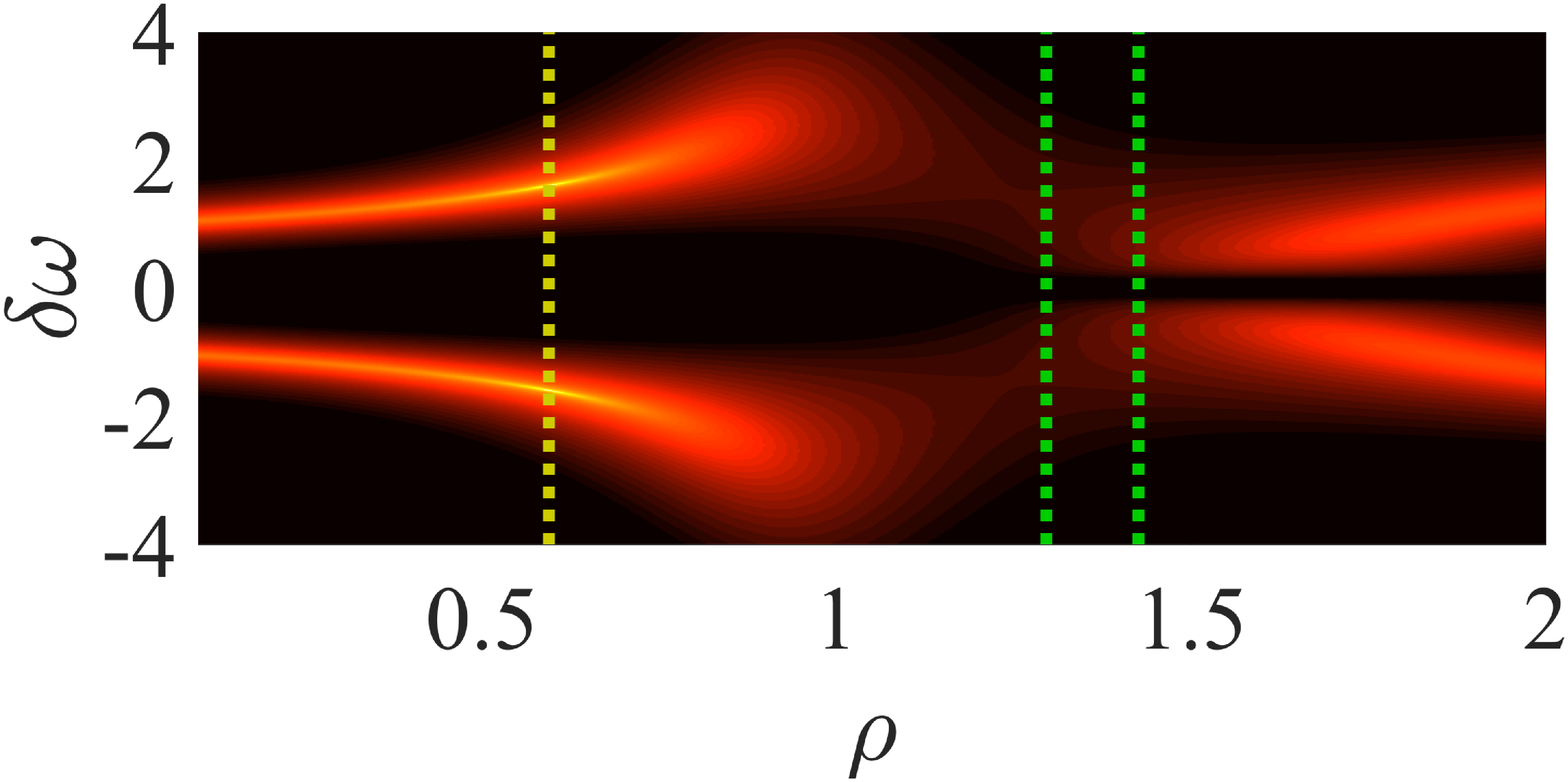}}}\\
    \caption{Eigenvalues (top, middle) and spectral line shape (bottom) of the phase-locked states for $\alpha=5$, $\Delta \neq 0$ and $P_1\neq P_2$. The vertical yellow and green dotted lines correspond to zero-crossings of the eigenvalue real part and exceptional points, respectively. (left) Phase-locked states for $\log\Lambda=-1.9$, $X_0=\sqrt{0.5}$ and $\theta=\pi$. The phase-locked states are stable between the first and the second as well as between the third and the fourth (left to right) vertical yellow dotted lines. No exceptional points exist in this case. (right) Phase-locked states for $\log \Lambda=1$, $X_0=\sqrt{0.5}$ and $\theta=\pi/2$. The phase-locked states are stable on the left side of the vertical yellow dotted line. The spectral line shape follows the imaginary part of the eigenvalues. Exceptional and bifurcation points are manifested in the spectral line shape through the emergence of side bands and intensity peaks, respectively.}
  \end{center}
\end{figure}

By linearizing the system (\ref{array}) around a fixed point corresponding to a phase-locked state we obtain the dynamical spectrum of the system in terms of the eigenvalues of its Jacobian ($\mathbf{J}$). The eigenvalues depend crucially on the asymmetry ($\rho$) and the phase-difference ($\theta$) of the phase-locked states and their trajectories in the parameter space have a rich and complex topology as shown in Figs. 4 and 5 for zero and non-zero  detuning, respectively. This topology dictates the bifurcations where the stability of the phase-locked states changes as the real part of an eigenvalues cross the zero axes (vertical yellow dotted lines) as well as where two or more eigenvalues coalesce as they meet at an exceptional point (vertical green dotted lines). Although, the zero-crossings of the real part of the spectrum corresponding to stability changes is manifested by the observable transition of the system either to another stable phase-locked state, or to a stable limit cycle, or to a chaotic state, the observable features that are related to the imaginary part of the eigenvalues and the existence of exceptional points are more elusive.\

In order to study the coherence properties of a specific phase-locked state we study the response of the linearized system under noise sources. The respective system has the form 
\begin{equation}
 \dot{\overrightarrow{\delta X}}=\mathbf{J}\overrightarrow{\delta X}+\overrightarrow{n} \label{linear_noise}
\end{equation}
where $\overrightarrow{\delta X}$ denotes small deviations around the phase-locked state $\overrightarrow{X}=(X_1,X_2,\theta,Z_1,Z_2)$, and $\overrightarrow{n}$ corresponds to amplitude, phase and carrier density fluctuations represented by zero-mean, delta-correlated stochastic signals. By taking the Fourier transform of Eq. (\ref{linear_noise}) we obtain the transfer function (matrix) of the system as
\begin{equation}
 \mathbf{H(\delta \omega)}=(i\delta \omega \mathbf{I}-\mathbf{J})^{-1}
\end{equation}
where $\delta \omega$ is the spectral component of the system response and $\mathbf{I}$ is the unit matrix.
The power spectral density $S_{XX}$ of the output of the linear system (\ref{linear_noise}) is then given by \cite{Papoulis_book} 
\begin{equation}
 \mathbf{S_{XX}}(\delta \omega)=\mathbf{H}(\delta \omega)\mathbf{H}^\dagger(\delta \omega)
\end{equation}
where the $\dagger$ denotes the Hermitian conjugate of the matrix. The spectral line shape of the phase-locked states in terms of the total optical power is given by $S_{X_1X_1}+S_{X_2X_2}$ and shown in Figs. 4 and 5 (bottom rows). It is clear that in all cases, the spectral line shape and its dependence on the specific phase-locked state (as a function of $\rho$) follows that of the imaginary part of the eigenvalues. The observable spectral signature of the exceptional points (vertical green dotted lines) corresponds to the emergence of side-bands and the observable spectral signature of the bifurcations at zero-crossings of the real part of the eigenvalues (vertical yellow dotted lines) corresponds to the appearance of points of maximum intensity. Thus, it is quite clear that the detuning and the asymmetric pumping determine the spectral line shape by controlling these points, resulting in a large set of qualitatively distinct spectral features. \

In summary, we have studied the fundamental active photonic dimer with the utilization of a realistic rate equation model. In contrast to simplified coupled mode equations models, the existence of spectral transitions and exceptional points is not restricted by PT-symmetry and zero detuning, suggesting a paradigm shift in non-Hermitian photonics. A plethora of dynamical features related to exceptional points and bifurcations has been discovered and their observable spectral signatures have been demonstrated. The freedom of the parameter selection in terms of detuning and more importantly of asymmetric pumping and the control of spectral characteristics suggest a versatile and multifunctional element for integrated active photonics and photonic metasurfaces.   

\section*{Acknowledgements}
This research is partly supported by two ORAU grants entitled "Taming Chimeras to Achieve the Superradiant Emitter" and ''Dissecting the Collective Dynamics of Arrays of Superconducting Circuits and Quantum Metamaterials'', funded by Nazarbayev University.

\end{document}